\documentclass[conference]{IEEEtran}

\usepackage[dvipsnames]{xcolor}

\usepackage[colorlinks=true,linkcolor=blue,breaklinks=True,citecolor=brown,urlcolor=blue]{hyperref}

\usepackage{dingbat}

\usepackage{adjustbox}
\usepackage[linesnumbered,algo2e,ruled]{algorithm2e}
\usepackage{multirow}
\usepackage{booktabs,subcaption,dcolumn}
\usepackage{tikz}
\usepackage{enumitem}
\setlist[itemize]{noitemsep, topsep=0pt}
\usepackage{tabularx}
\usepackage[normalem]{ulem}

\usepackage{pifont}
\usepackage{svg}
\usepackage{soul}
\usepackage[font=footnotesize]{caption}
\usepackage[noadjust]{cite}

\usepackage{amsmath}
\usepackage{amsfonts, amssymb}
\usepackage{colortbl}
\usepackage{tablefootnote}
\usepackage[most]{tcolorbox}
\usepackage[sort&compress, numbers]{natbib}

\AtBeginDocument{%
  \providecommand\BibTeX{{%
    \normalfont B\kern-0.5em{\scshape i\kern-0.25em b}\kern-0.8em\TeX}}}

\newcommand{\cmark}{{\color{ForestGreen}{\ding{51}}\xspace}}
\newcommand{\xmark}{{\color{RubineRed}{\ding{55}}\xspace}}
\newcolumntype{?}{!{\vrule width 1.5pt}}

\newcommand{\companyname}{{\small \textit{Sigma}}\xspace}

\newcommand{\tablength}{12cm}
\newcommand{\tabblength}{10cm}

\newcommand{\textbox}[1]{
    \noindent\fbox{%
        \parbox{0.97\columnwidth}{%
            {#1}
        }%
    }
}

\newtcolorbox{cooltextbox}[1][]{%
    colback=black!5,
    colframe=black!5,
    notitle,
    sharp corners,
    borderline west={0pt}{0pt}{red!80!black},
    enhanced,
    breakable,
    left=0pt,
    right=0pt,
    top=0pt,
    bottom=0pt
    }

\newcommand\smamath[1]{{\small $#1$}}

\newcommand\ftmath[1]{{\footnotesize $#1$}}

\newcommand\revision[1]{%
  \bgroup
  \hskip0pt\color{blue!80!black}%
  #1%
  \egroup
}

\begin{document}

\title{``Do Users fall for Real Adversarial Phishing?''\\
Investigating the Human response to Evasive Webpages}

\author{

\IEEEauthorblockN{{Ajka Draganovic\IEEEauthorrefmark{1}, Savino Dambra\IEEEauthorrefmark{5}, Javier Aldana Iuit\IEEEauthorrefmark{4}, Kevin Roundy\IEEEauthorrefmark{5}}, Giovanni Apruzzese\IEEEauthorrefmark{1}\\}
\IEEEauthorblockA{{ 
\IEEEauthorrefmark{1}\textit{University of Liechtenstein},
\IEEEauthorrefmark{5}\textit{Norton Research Group},
\IEEEauthorrefmark{4}\textit{Avast Software}}
\\
{\small \{name.surname\}@\{uni.li\IEEEauthorrefmark{1}, gendigital.com\IEEEauthorrefmark{4}\IEEEauthorrefmark{5}\},
}}}

\pagestyle{plain}
\maketitle

\begin{abstract}

Phishing websites are everywhere, and countermeasures based on static blocklists cannot cope with such a threat. To address this problem, state-of-the-art solutions entail the application of machine learning (ML) to detect phishing websites by checking if they visually resemble webpages of well-known brands. These techniques have achieved promising results in research and, consequently, some security companies began to deploy them also in their phishing detection systems (PDS). However, ML methods are not perfect and some samples are bound to bypass even production-grade PDS. 

In this paper, we scrutinize whether \textit{genuine phishing websites} that evade \textit{commercial ML-based PDS} represent a problem ``in reality''. Although nobody likes landing on a phishing webpage, a false negative may not lead to serious consequences if the users (i.e., the actual target of phishing) can recognize that ``something is phishy''. 
Practically, we carry out the first user-study (N=126) wherein we assess whether unsuspecting users (having diverse backgrounds) are deceived by ``adversarial'' phishing webpages that evaded a real PDS. We found that some well-crafted adversarial webpages can trick most participants (even IT experts), albeit others are easily recognized by most users.
Our study is relevant for practitioners, since it allows prioritizing phishing webpages that simultaneously fool (i) machines and (ii) humans---i.e., their intended targets.

\end{abstract}

\pagestyle{plain}
\maketitle

\section{Introduction}
\label{sec:introduction}
\noindent
The battle against phishing is still ongoing~\cite{proofpoint2022phish}, despite decades of efforts aimed at countering this threat~\cite{khonji2013phishing, biggio2018wild}. According to the FBI, phishing is the leading form of cyber-crime~\cite{fbi2021icr}, and its proliferation is constantly increasing~\cite{apwg2022}.

\begin{figure}[!htbp]
    \centering
    \includegraphics[width=\columnwidth]{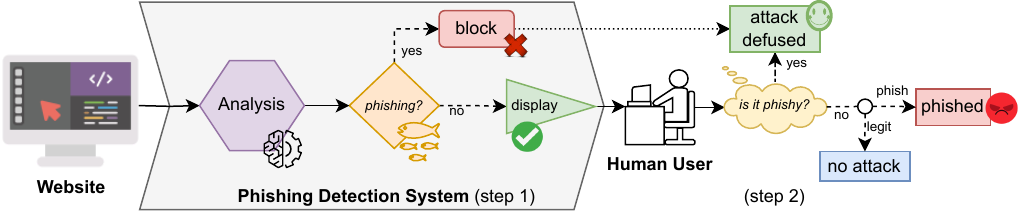}
    \caption{Scenario: phishing detection is a two-step decision process.}
    \label{fig:scenario}
\end{figure}

Phishing \textit{websites} are among the most common vectors used by adversaries to carry out phishing attacks~\cite{baki2022sixteen}. After deploying their phishing hooks ``in the wild,'' attackers try to lure their victims (through, e.g., social engineering) to such malicious webpages---intent to steal their private data, or compromise their IT systems. Countermeasures to phishing can fall in two categories: \textit{human-centered} (e.g., phishing awareness training~\cite{orunsolu2018users}), which aim at improving the ability of humans to avoid phishing traps; and \textit{machine-centered} (e.g., phishing website detectors~\cite{abdelnabi2020visualphishnet}), which aim at preventing the human user from landing on a phishing trap in the first place. As a matter of fact, the fight against phishing can be seen as a \textbf{two-step decision process}, which we illustrate in Fig.~\ref{fig:scenario}. After a user is brought to any given website, a phishing detection system (PDS) quickly analyzes the website (e.g., by checking some blocklists or using heuristics~\cite{bell2020analysis}): if the PDS determines the website to be phishing, then the webpage is not displayed to the user (who might be shown a warning/alert); otherwise, the browser renders the webpage. Of course, no issue arises if the webpage is benign. However, if the webpage is malicious, the decision is now up to the user: if they can recognize the page as phishing, then the attack is defused; otherwise, the user (i.e., its data or device) may be ``caught''.

Unfortunately, \textit{operational} PDS are tweaked to minimize the rate of false alarms, which leads to a significant number of phishing websites to \textit{evade} their detection (a security company had over 9k ``evasions'' in just one month~\cite{apruzzese2023real}).\footnote{This is why PDS have been relying on blocklists for a long time~\cite{oest2020phishtime}, albeit state-of-the-art PDS now also leverage artificial intelligence to provide an additional layer of defense~\cite{divakaran2022phishing, apruzzese2023real}. See §\ref{ssec:pwd} for background.} Given the brittleness of existing anti-phishing schemes, it is paramount to improve the users' ability to autonomously recognize phishing websites. However, according to a recent Proofpoint's report~\cite{proofpoint2022phish}, more than 33\% companies do not have any training program; and, among those which do provide such training, more than 50\% do so via simulations.

Amidst this chaos, we observe that there is a \textbf{mismatch between research efforts that focus on human- or machine-centered} solutions. In particular, despite phishing detection being a two-step decision process, prior work only focused on either one of these steps. For example, papers that propose novel PDS tend to overlook how humans respond to those webpages that bypassed the proposed PDS; whereas papers that focus on the human perception of phishing websites do not consider webpages that evaded operational PDS (we discuss related work in §\ref{ssec:related}). Such a disconnection is problematic: a PDS whose false negatives can trick all end users (i.e., the true target of phishing) is not reliable; whereas carrying out phishing assessments using webpages that can be trivially blocked by PDS may not be the best way to invest resources. 

In this paper, we seek to bridge the gap between these two complementary approaches against phishing. To this end, we reach out to a security company that develops anti-phishing schemes, and obtain a set of ``adversarial phishing webpages''~(AW) that evaded their operational PDS powered by deep learning (§\ref{ssec:data}). Then, we carry out a user-study (N=126) in which we ask participants (who were not primed in any way) to figure out whether such AW resemble legitimate websites or not (§\ref{ssec:questionnaire}). We also inquire potential explanations for their skepticism (if any). We analyze our results both quantitatively (§\ref{sec:results}) and qualitatively (§\ref{sec:explanations}). Our findings reveal that while ``poorly crafted'' AW can be easily spotted by end-users, others can deceive (almost) all of our participants.

\textsc{\textbf{Contribution.}}
To bridge the gap between human- and machine-centered anti-phishing schemes, we:
\begin{itemize}[leftmargin=0.4cm]
    \item Carry out the \textit{first} user-study elucidating the response of humans to \textit{real} phishing webpages that evaded a \textit{real} phishing detection system based on deep learning.
    
    \item Validate our findings quantitatively---via \textit{statistical} tests (§\ref{ssec:statistical}); and qualitatively---alongside \textit{practitioners} (§\ref{ssec:interpretation}); and derive \textit{recommendations} for research (§\ref{ssec:recommendations}).
    
    \item Provide \textit{practical insights} on operational PDS (§\ref{ssec:data}) and on how to improve them (we share our phishing data~\cite{repository}); 
\end{itemize}
This study can spearhead future work aimed at improving PDS, i.e., by identifying the AW that deceive most users, and then fixing PDS so that such AW are not misclassified.
\section{Background and Motivation}
\label{sec:background}
\noindent
We focus on phishing \textit{websites}. Other forms of phishing (e.g., email~\cite{ampel2023benchmarking, kersten2022investigating,sheng2010falls}) are outside our scope---albeit our findings can apply to these (if they envision luring a victim to a website). 

To allow a complete understanding of the problem tackled by our paper (§\ref{ssec:problem}), we summarize the landscape of phishing website detection (§\ref{ssec:pwd}), and then position our paper within existing literature on the human perception of phishing (§\ref{ssec:related}).

\subsection{Phishing Website Detection}
\label{ssec:pwd}
\noindent
The first line of defense against phishing entails \textit{automated} detection schemes~\cite{Corona:Deltaphish}. The goal is analysing a given website to determine whether it is malicious (or not) and, if so, prevent its webpage from being displayed to an end-user. 

\textbf{Rule-driven detection.}
The most popular way to fight phishing websites is through \textit{blocklists}~\cite{oest2020phishtime, kondracki2021catching, feal2021blocklist}: by checking if an URL (or part of it) is included in a pre-defined set of malicious URLs (or domains), it is possible to precisely (and quickly) identify phishing websites. Detection mechanisms based on blocklists are widespread in modern browsers (e.g., Google Safe Browsing~\cite{bell2020analysis}), and are appreciated due to their near-zero false positive rate. Unfortunately, even though such mechanisms are kept up-to-date with new malicious entries, these tools are useless against ``novel'' phishing websites~\cite{tian2018needle}.

\textbf{Data-driven detection.} 
To protect users against phishing websites that are not included in any blocklist, abundant research efforts proposed data-driven solutions based on heuristics. For instance, Zhang et al.~\cite{zhang2007cantina} identified some patterns commonly associated to phishing, and used these to discriminate benign from phishing websites. Similar detection techniques also encompass machine learning (ML) methods~\cite{apruzzese2022role}. For instance, Mohammad et al.~\cite{mohammad2014predicting} proposed a set of features (extracted from the URL and the HTML of a webpage) that could be used to develop a ML-based detector (after undergoing a proper training phase), whereas Cui et al.~\cite{cui2020semanticphish} use ML to infer malicious domains typically associated to phishing. A complementary data-driven approach against phishing entails using the \textit{visual similarity}. Early works date back to 2006~\cite{fu2006detecting}, and some even leverage ML (e.g.,~\cite{Corona:Deltaphish}). More recently, due to the never-ending advancement of deep learning~(DL), detection methods reliant on visual cues attracted much attention in research~\cite{van2021combining,abdelnabi2020visualphishnet,lin2021phishpedia,liu2022inferring}. However, despite abundant scientific literature, these proposals suffer from a significant drawback: the ``high'' false positive rate---which impairs the end-user experience. Consequently, the integration of ML/DL into operational phishing detection systems (PDS) proceeds at a slow pace---but it is happening~\cite{divakaran2022phishing}.

\textbf{Adversarial phishing.} 
Besides having to deal with false positives, \textit{real} PDS must face another issue: the ``false negatives'' that stem from adversaries who deliberately want to evade the PDS~\cite{biggio2018wild}. Indeed, abundant evidence suggests that even data-driven solutions cannot ``catch-all-phish''. Liang et al.~\cite{liang2016cracking} cracked Google's page filter in 2016. More recently, security enthusiasts bypassed the ML-based detector of a popular anti-phishing evasion competition~\cite{gao2023evading}; Apruzzese et al.~\cite{spacephish2022} showed that state-of-the-art detectors (analyzing either the URL/HTML) can be fooled via cheap perturbations, whereas Lee et al.~\cite{lee2023attacking} evaded logo-based detectors proposed in research with imperceptible visual changes. Finally, a recent work~\cite{apruzzese2023real} showed that even \textit{commercial-grade} PDS that uses DL for visual similarity exhibits thousands of false negatives every month---some of which due to ``perturbations'' that are easily recognizable by humans.

\subsection{Problem Statement (Focus of the Paper)}
\label{ssec:problem}
\noindent
Detecting phishing websites is tough, and false positives/negatives are bound to occur. In reality, however, practitioners are vexed by a dilemma: ``what to prioritize?''~\cite{apruzzese2023real}. The answer should be driven by the \textbf{perspective of the end-user}---the true target of phishing.

Nobody wants their browsing activities to be frequently interrupted by inaccurate blocking mechanisms (i.e., false positives). However, by turning the attention to the \textit{false negatives} of a PDS (which lead to displaying a malicious page), there are two cases:
\begin{enumerate}
    \item the user \textit{recognizes} the page as phishing. Despite being an annoyance (``yet another phishing website!''), the consequences of such a misclassification are mild---the user will simply close the webpage and resume their activities.
    \item the user \textit{does not} recognize the page as phishing. This is a serious problem, since it may lead to the phishing attack being successful---causing a much greater loss (in terms of time, finance, or privacy~\cite{hutchings2017online}) to the user.
\end{enumerate}
In this paper, we are inspired by such ``dual nature'' of \textbf{false negatives} in the context of phishing website detection. We seek to scrutinize the response of humans to those ``adversarial webpages'' that evaded a ML/DL-based PDS. 
From a practical viewpoint, the findings of our study can assist practitioners in optimizing their (limited) resources, e.g., by placing more emphasis on those pages that can deceive users.
To the best of our knowledge, we are the first to investigate this problem.

\subsection{Related Work (User Studies)}
\label{ssec:related}
\noindent
Let us discuss our study in light of existing literature.

\textbf{Technical papers.}
Abundant research on phishing website detection entail ``technical papers'', typically proposing either a defensive solution that improves the state-of-the-art (e.g.,~\cite{zhang2022m,apruzzese2022mitigating}); or a new attack that bypasses existing anti-phishing schemes (e.g.,~\cite{spacephish2022}). Unfortunately, most such papers overlook the user perspective; with two notable exceptions: 
{\small \textbf{(1)}}~Abdelnabi et al.~\cite{abdelnabi2020visualphishnet}, after proposing a novel detection technique, carried out a user-study asking participants to ``evaluate how trustworthy [misclassified webpages] seem based on visual similarity''. From a realistic viewpoint, however, such a user-study has two limitations: {\small \textit{(i)}}~it is based on the output of a research proposal---and not on a \textit{real} PDS; and {\small \textit{(ii)}}~the phrasing of the question \textit{primes} the users---who are more likely to be suspicious of a webpage, and which inevitably leads to biased results. 
{\small \textbf{(2)}}~Lee et al.~\cite{lee2023attacking} propose a new means to evade PDS based on logo-identification, and then carry out a user-study wherein humans are asked to identify how similar ``adversarial logos'' are to the original logos of well-known brands. However, besides being also based on a research proposal, this user-study only focuses on the logo---which is only a small part of a real webpage, and is hence inappropriate to derive whether the user would be truly fooled by the corresponding webpage.

\textbf{Human-centered papers.}
Differently from technical papers, another branch of research specifically focuses on investigating the human factor in phishing (for, e.g., educational training campaigns~\cite{baki2022sixteen}). Such ``human-centered'' papers are closer to our work. However, none of these papers investigate (neither explicitly nor implicitly) the specific problem tackled in our research. To position our paper within related works, we carry out an extensive literature review\footnote{We perform the literature review between November 2022 and June 2023. During this time-frame, two authors manually queried popular scientific repositories for user-studies on the perception of humans to phishing websites. The authors frequently met and discussed their individual findings across various meetings to derive a consensus. We omitted three works (\cite{iuga2016baiting, lastdrager2017effective, orunsolu2018users} because they focus on websites \textit{and emails}---which are outside our scope.} wherein we scrutinize each related work according to four criteria:
\begin{itemize}[leftmargin=0.5cm]
    \item \textit{Deployed ML misclassifications}: did the phishing webpages in the user-study evade a real ML-based detector?
    \item \textit{No priming}: were the participants kept in the dark about the study being about phishing? (otherwise, it can bias results)
    \item \textit{Real phishing}: were the phishing webpages taken from ``the wild web''? (perhaps such pages were created in a lab)
    \item \textit{IT expertise}: was the IT expertise accounted for? (experts in IT may respond differently than amateurs)
\end{itemize}
These criteria allow one to assess the ``realistic value'' of the findings of each prior user-study. We summarize our literature review in Table~\ref{tab:related}, in which we also report the amount of participants included in the user-study. 

\vspace{-1mm}

\begin{table}[!htbp]
\caption{User-studies on the human perception of phishing websites. A ``?'' denotes works for which we could not find any information.}
\label{tab:related}
\vspace{-2mm}

\resizebox{1\columnwidth}{!}{

\begin{tabular}{ccccccc}
\toprule
\textbf{Paper} & \textbf{Year} & 
\begin{tabular}{c}
     \textbf{Sample}\\
     \textbf{size}
\end{tabular} &
\begin{tabular}{c}
    \textbf{Deployed ML} \\
    \textbf{missclass.}
\end{tabular} &
\begin{tabular}{c}
    \textbf{No} \\
    \textbf{Priming}
\end{tabular} &
\begin{tabular}{c}
    \textbf{Real} \\
    \textbf{Phishing}
\end{tabular} &
\begin{tabular}{c}
    \textbf{IT} \\
    \textbf{Expertise}
\end{tabular} \\
\midrule
 Dhamija~\cite{dhamija2006phishing} &  2006 &  22 &  \xmark &  \xmark &  \xmark & \cmark \\
 Tsow~\cite{tsow2007deceit} &  2007 &  398 &  \xmark &  \xmark &  \xmark &  ? \\
 
 Sheng~\cite{sheng2007anti} & 2007 & 42 & \xmark & \xmark & \cmark & \cmark \\
 
 Jakobsson~\cite{jakobsson2007human} &  2007 &  400 &  \xmark &  \xmark & ? &  \xmark \\

 Herzberg~\cite{herzberg2008security} & 2008 & 23 & \xmark & \xmark & \xmark & \xmark \\ 
 
 Alnajim~\cite{alnajim2009anti} & 2009 & 36 & \xmark & \cmark & \cmark & \cmark \\
 
 Kumaraguru~\cite{kumaraguru2010teaching}&  2010 &  28 &  \xmark &  \xmark &  \cmark &  \cmark \\

 Yang~\cite{yang2012building} & 2012 & 62 & \xmark & \xmark & \xmark & \xmark \\
 
 Asanka~\cite{asanka2013can} & 2013 & 40 & \xmark & \xmark & \xmark & \cmark \\
 
 Purkait~\cite{purkait2014empirical}&  2014 &  621 &  \xmark &  \xmark &  \xmark &  \cmark \\

 Scott~\cite{scott2014assessing} & 2014 & 66 & \xmark & \xmark & \xmark & \xmark \\

 Alsharnouby~\cite{alsharnouby2015phishing} &  2015 &  21 &  \xmark & ? &  \xmark &  \cmark \\
 
 Kunz~\cite{kunz2016nophish} & 2016 & 32 & \xmark & \xmark & ? & \xmark \\

 Arachilage~\cite{arachchilage2016phishing} & 2016 & 20 & \xmark & \xmark & \cmark & \cmark\\
 

 Xiong~\cite{xiong2017domain} & 2017 & 320 & \xmark & \cmark & \xmark & \cmark \\

 Moreno~\cite{moreno2017fishing} & 2017 & 175 & \xmark & \xmark & \xmark & \xmark \\
 
 
 Gopavaram~\cite{gopavaram2021cross} &  2021 &  250 &   \xmark &  \xmark &  \xmark &  \cmark \\
\midrule
 \textbf{This Paper} &  \textbf{2023} &  \textbf{126} &  \cmark &  \cmark &  \cmark &  \cmark \\
\bottomrule
\end{tabular}
}

\end{table}

\textbf{Considerations.}
By observing Table~\ref{tab:related}, we can see that most papers do not meet all such criteria. In particular, no paper considered ``Deployed ML misclassifications'': this is not surprising, given that most PDS are closed-source (hence it is not known whether they integrate some ML or not) and deployment of ML in cyber security proceeds at a slow-pace~\cite{apruzzese2022role}. We also mention that many studies tend to prime their participants, which may not represent a realistic scenario (if a user ``expects'' to encounter phishing, they are less likely to fall for it in the first place~\cite{bullee2020effective}). 

\vspace{1mm}

\textbox{\textbf{Summary:} prior user-studies do not allow investigating the \textit{real} response of humans to \textit{real} phishing webpages that evaded a \textit{real} ML-based phishing detection system.}

\noindent
\textbf{Why ML?} 
We are aware that many data-driven methods exist to fight phishing (§\ref{ssec:pwd}). We focus on ML-based PDS (using visual similarity) due to their emerging deployment in the real-world~\cite{apruzzese2023real}, in an attempt to raise the awareness on the concrete issues of these schemes \textit{before} they become widespread---given how easily they can be bypassed~\cite{lee2023attacking, apruzzese2023real}.
\section{Research Method}
\label{sec:method}
\noindent
Our study revolves around a central research question (RQ): ``\textit{How do users respond to phishing webpages that evaded a ML-based PDS?}''. To answer our RQ, we first obtain a set of \textit{adversarial} phishing webpages (§\ref{ssec:data}), and then devise a questionnaire (§\ref{ssec:questionnaire}) aimed at assessing the \textit{awareness} of users (§\ref{ssec:sample}) to such webpages.

 \begin{figure*}[!htbp]
    \centering
    \includegraphics[width=1.5\columnwidth]{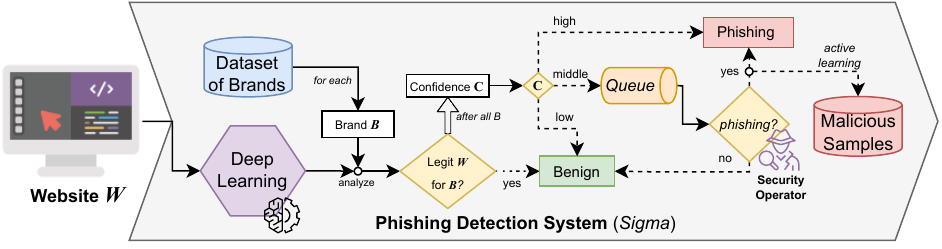}
    \caption{The architecture of the PDS deployed by \textit{Sigma}, used as basis for the phishing examples to include in our user-study.}
    \label{fig:sigma}
\end{figure*}

\subsection{Data Source}
\label{ssec:data}

\textbf{Highlight.} A pivotal characteristic of our research is that we use data pertaining to a \textit{real system}. Indeed, the webpages included in our questionnaire represent \textit{real phishing webpages} that are encountered ``in the wild'' by \textit{real users}, and which manage to bypass one of the components of a \textit{commercial-grade} phishing website detector reliant on deep learning.
We reached out to a large security company (which we refer to as ``\companyname'') whose services entail phishing website protection. In particular, \companyname employs diverse defensive mechanisms that work in tandem to minimize the chance that users fall for potential phishing ``hooks''. Among these, \companyname also provides a detector that leverages state-of-the-art techniques based on visual similarity (e.g.,~\cite{abdelnabi2020visualphishnet,lin2021phishpedia}) to identify phishing websites. 

 \textbf{Phishing Detection System.} The PDS used by \companyname seeks to identify phishing websites attempting to impersonate known brands (e.g., PayPal). Such a PDS (a schematic is in Fig.~\ref{fig:sigma}) processes diverse streams of URLs for which it tries to infer whether the corresponding website $W$ is legitimate or not. This is done by using various DL models to compute the visual similarity between $W$ and the entire dataset of websites associated to the brands tracked by \companyname (i.e., a given brand may have more than one website): if $W$ is found to be ``identical'' to any website included in such dataset, then it is flagged as benign (and, potentially, further analysed by other mechanisms~\cite{lin2021phishpedia}). Otherwise, the decision of the PDS depends on the confidence $C$ computed by the DL models for each website in the dataset of brands. Specifically, the PDS takes the top value of $C$ (which is related to a specific brand $B$), and then compares such $C$ against a given threshold. If $C$ is ``high'', then $W$ is considered as phishing (i.e., $W$ is trying to impersonate $B$); if $C$ is ``low'', then $W$ is considered as benign (i.e., $W$ is very different from $B$, and it may be a website of an unknown brand); if $C$ is in-between, then $W$ is not marked straight away as malicious (to avoid raising potential false positives), and is then put in a dedicated queue meant to be manually inspected by security operators. This is because \companyname seeks to improve their services by having samples that are ``difficult to classify'' to be used as basis to develop more robust detectors (i.e., an active-learning\footnote{\textit{Sigma} aims to improve their PDS by having $W$ be checked also against ``known malicious examples'' -- triggering an immediate phishing response.} approach~\cite{apruzzese2022role}).

\textbf{Adversarial webpages.} After finding an agreement with \companyname (which required an NDA), we were given hundreds of webpages (as screenshots---in full HD resolution and taken in late 2022) which fell in-between the ``high'' and ``low'' confidence. Hence, such webpages had undergone manual verification by security analysts who verified that all such webpages were \textit{phishing} websites disguised as benign webpages---thereby representing ``false negatives''. Interestingly, most of these phishing webpages were ``poorly crafted'': most humans would probably be able to suspect that ``something is phishy''. However, there were also cases in which the webpages exhibited a remarkable similarity with the webpage they were attempting to mimic (we provide some examples in Figs.~\ref{fig:question} and~\ref{fig:screenshots}). For the sake of our RQ and given the exploratory nature of our study, we considered samples from both categories. In the remainder, we will use the term ``adversarial webpage'' (AW) to denote a \textit{screenshot} of a phishing webpage that bypassed the stricter threshold of the PDS deployed by \companyname and that was hence displayed to end users.

\subsection{Questionnaire}
\label{ssec:questionnaire}

\textbf{Goals.} After receiving the AW from \companyname, we devised the questionnaire used to answer our RQ. In doing so, we adhered to the following design goals ($\Rightarrow$ is the motivation):
\begin{itemize}
    \item \textit{Heterogeneous sample}: anybody was eligible to participate in our user study. 
    $\Rightarrow$~Since we consider AW ``in the wild'', anyone can land upon them. Hence, given that we are the first to conduct such a study, we follow an exploratory approach and do not set any constraint in terms of, e.g., technical expertise. We will, however, ask participants to provide some background information to enable fine-grained analyses.
    \item \textit{No priming}: participants should not be aware that the questionnaire is related to phishing. $\Rightarrow$~The main reason why phishing attacks succeed is that users are distracted, or do not suspect that a given webpage may be malicious~\cite{ferreira2015analysis}. To investigate the real effectiveness of AW, we will not mention any term that may alert our participants.
    \item \textit{Brand knowledge}: we ensure that our participants are familiar with the (legitimate) websites ``mimiced'' by our AW. 
    $\Rightarrow$~To provide a significant answer to our RQ, we must focus on users who ``can'' be phished by a given AW (and inquire about their knowledge of IT).
\end{itemize}
Finally, our research is mostly tailored for Europe (due to \companyname's main location). Hence, our participants and questionnaire are going to reflect this side of the World. The list of the brands we considered (and supporting evidence) is in Table~\ref{tab:brands}.

\begin{table}[!htbp]
\caption{Brands included in our questionnaire.}
\label{tab:brands}

\resizebox{1\columnwidth}{!}{

\begin{tabular}{c|l|l}
\toprule
 \textbf{Brand} & \textbf{Category} & \textbf{Reason (and source)}  \\
\midrule

Netflix & \multicolumn{1}{p{1.5cm}|}{Video Streaming} & \multicolumn{1}{p{\tabblength}}{In Q4 2022, the Europe, Middle East, and Africa  demonstrated the highest concentration of paying customers for NetFlix (over 76M are from Europe)~\cite{netflixStats}.} \\
\hline

Amazon & eShop & \multicolumn{1}{p{\tabblength}}{Amazon operates in eight European countries (Germany being the most active, with €32B in net sales in 2021)~\cite{amazonStats}.}\\\hline

Zalando & eShop & 
\multicolumn{1}{p{\tabblength}}{Zalando is a prominent fashion and lifestyle platform in Europe, experienced a 6\% increase in active customers, surpassing 51M individuals in 2022~\cite{zalandoStats}.}\\

\hline
Airbnb & Travel & \multicolumn{1}{p{\tabblength}}{With 1.34M hosts, Europe is the largest AirBnB community worldwide~\cite{airbnbStats}.}\\\hline

Google & \multicolumn{1}{p{1.3cm}|}{Information and Email} & \multicolumn{1}{p{\tabblength}}{Google garners 89.3 billion monthly visits, while its email service, Gmail, enjoys widespread adoption across multiple European countries~\cite{gmailStats,googleStats}.}\\
\hline

Instagram & \multicolumn{1}{p{1.3cm}|}{Social Network} & \multicolumn{1}{p{\tabblength}}{In 2022, Europe stands as the second largest community of Instagram users, comprising an impressive population of 338 million individuals~\cite{instagramStats}.} \\ \hline

Facebook & \multicolumn{1}{p{1.3cm}|}{Social Network} & \multicolumn{1}{p{\tabblength}}{According to Meta, in Q4 2022 Facebook recorded 411M monthly active users in Europe (4 more than in Q3)~\cite{facebookStats}.} \\ \hline

LinkedIn & \multicolumn{1}{p{1.3cm}|}{Social Network} & \multicolumn{1}{p{\tabblength}}{LinkedIn, encompassing an extensive user base of nearly 1B individuals worldwide, has garnered traction within Europe, boasting a count of 242M users~\cite{linkedinStats}.} \\

\hline
PayPal & Banking & \multicolumn{1}{p{\tabblength}}{PayPal revealed that its user base in Europe amounts to 35M individuals~\cite{paypalStats}.} \\
\hline
Uber & Mobility &
\multicolumn{1}{p{\tabblength}}{Uber maintains a presence in a significant number of European countries~\cite{uberStats}.} \\
\hline
Yahoo & \multicolumn{1}{p{1.3cm}|}{Information and Email} & \multicolumn{1}{p{\tabblength}}{Based on the Alexa rankings, Yahoo.com attains the eleventh position among the most frequently accessed websites on a global scale~\cite{yahooStats}.}\\\hline
Twitter & \multicolumn{1}{p{1.3cm}|}{Social Network} & \multicolumn{1}{p{\tabblength}}{Despite being less popular than in the previous decade, Twitter is still popular in Europe (70M active users in 2023)~\cite{twitterStats}.} \\

\bottomrule
\end{tabular}
}

\end{table}

\begin{table*}[!htbp]
\caption{Sequence of screenshots in our questionnaire, and their difficulty level. The number points to the image (hosted in our repo~\cite{repository}).}
\label{tab:questionnaire}
\centering
\resizebox{1.8\columnwidth}{!}{

\begin{tabular}{c|c|c|l}
\toprule
\textbf{\#} & \textbf{Brand} & \textbf{Difficulty} & \textbf{Comment}\\ 

\midrule
\href{https://github.com/hihey54/eCrime23_realAdversarialPhish/blob/main/screenshots/1.jpeg}{1} & Instagram & \textit{Hard} & \multicolumn{1}{p{\tablength}}{Resembles the legitimate login page, with the sole distinction being the footer's style.} \\ 
\hline
\href{https://github.com/hihey54/eCrime23_realAdversarialPhish/blob/main/screenshots/2.jpeg}{2} & Facebook & \textit{Moderate} & \multicolumn{1}{p{\tablength}}{Appears similar to the authentic version; however, suspicion may arise due to the multiple profiles that have recently logged in from the same device (specifically, six different profiles).} \\ 
\hline
\href{https://github.com/hihey54/eCrime23_realAdversarialPhish/blob/main/screenshots/3.jpeg}{3} & Facebook & \textit{Hard} & \multicolumn{1}{p{\tablength}}{Closely resembles the original, with the sole exception of a missing footer.} \\ 
\hline
\href{https://github.com/hihey54/eCrime23_realAdversarialPhish/blob/main/screenshots/4.jpeg}{4} & Instagram & \textit{Hard} & \multicolumn{1}{p{\tablength}}{Extremely challenging to distinguish, as it perfectly mirrors the original.} \\ 
\hline
\href{https://github.com/hihey54/eCrime23_realAdversarialPhish/blob/main/screenshots/5.jpeg}{5} & PayPal & \textit{Hard} & \multicolumn{1}{p{\tablength}}{Resembles the authentic site very closely.} \\
\hline
\href{https://github.com/hihey54/eCrime23_realAdversarialPhish/blob/main/screenshots/6.jpeg}{6} & Google & \textit{Hard} & \multicolumn{1}{p{\tablength}}{Resembles the authentic site very closely.} \\ 
\hline
\href{https://github.com/hihey54/eCrime23_realAdversarialPhish/blob/main/screenshots/7.jpeg}{7} & Amazon & \textit{Moderate} & \multicolumn{1}{p{\tablength}}{Resembles the authentic site very closely, but some elements have a different style.} \\ 
\hline
\rowcolor{green!20}\href{https://github.com/hihey54/eCrime23_realAdversarialPhish/blob/main/screenshots/8.jpeg}{8} & Airbnb & --- & \multicolumn{1}{p{\tablength}}{It is the legitimate website.} \\ 
\hline
\rowcolor{green!20}\href{https://github.com/hihey54/eCrime23_realAdversarialPhish/blob/main/screenshots/9.jpeg}{9} & Zalando & --- & \multicolumn{1}{p{\tablength}}{It is the legitimate website.} \\ 
\hline
\href{https://github.com/hihey54/eCrime23_realAdversarialPhish/blob/main/screenshots/10.jpeg}{10} & Netflix & \textit{Moderate} & \multicolumn{1}{p{\tablength}}{The website's header and logo may induce suspicion due to their uncharacteristic design.} \\
\hline
\href{https://github.com/hihey54/eCrime23_realAdversarialPhish/blob/main/screenshots/11.jpeg}{11} & Yahoo & \textit{Moderate} & \multicolumn{1}{p{\tablength}}{Resembles the authentic site, but some elements are stretched.} \\ 
\hline
\href{https://github.com/hihey54/eCrime23_realAdversarialPhish/blob/main/screenshots/12.jpeg}{12} & Yahoo & \textit{Hard} & \multicolumn{1}{p{\tablength}}{Resembles the authentic site very closely.} \\ 
\hline
\href{https://github.com/hihey54/eCrime23_realAdversarialPhish/blob/main/screenshots/13.jpeg}{13} & Netflix & \textit{Easy} & \multicolumn{1}{p{\tablength}}{The font style noticeably deviates from the one typically used.} \\ 
\hline
\href{https://github.com/hihey54/eCrime23_realAdversarialPhish/blob/main/screenshots/14.jpeg}{14} & Uber & \textit{Easy} & \multicolumn{1}{p{\tablength}}{The appearance of Uber's sign-in page notably diverges from the expected layout.} \\ 
\hline
\href{https://github.com/hihey54/eCrime23_realAdversarialPhish/blob/main/screenshots/15.jpeg}{15} & PayPal & \textit{Moderate} & \multicolumn{1}{p{\tablength}}{The background color of the input fields clashes with the overall design aesthetic of the website.} \\
\hline
\href{https://github.com/hihey54/eCrime23_realAdversarialPhish/blob/main/screenshots/16.jpeg}{16} & Uber & \textit{Easy} & \multicolumn{1}{p{\tablength}}{The appearance suggests it might be an outdated version of Uber.} \\ 
\hline
\href{https://github.com/hihey54/eCrime23_realAdversarialPhish/blob/main/screenshots/17.jpeg}{17} & LinkedIn & \textit{Easy} & \multicolumn{1}{p{\tablength}}{The font style significantly deviates from what one would expect on a professional website, disrupting its overall look and feel.} \\ 
\hline
\href{https://github.com/hihey54/eCrime23_realAdversarialPhish/blob/main/screenshots/18.jpeg}{18} & Netflix & \textit{Very easy} & \multicolumn{1}{p{\tablength}}{No resemblance to the original sign-up page, with a starkly contrasting and distinctive styling.} \\ 
\hline
\href{https://github.com/hihey54/eCrime23_realAdversarialPhish/blob/main/screenshots/19.jpeg}{19} & Twitter & \textit{Moderate} & \multicolumn{1}{p{\tablength}}{It gives the impression of being an older version of Twitter, which could still potentially elicit trust from unfamiliar users.} \\ 
\hline
\href{https://github.com/hihey54/eCrime23_realAdversarialPhish/blob/main/screenshots/20.jpeg}{20} & Amazon & \textit{Moderate} & \multicolumn{1}{p{\tablength}}{While it bears a striking resemblance, participants might grow suspicious due to the button on the page appearing incongruous with the overall design.} \\ 
\bottomrule
\end{tabular}
}

\end{table*}

\begin{figure}[!htbp]
    \centering
    \includegraphics[width=0.98\columnwidth]{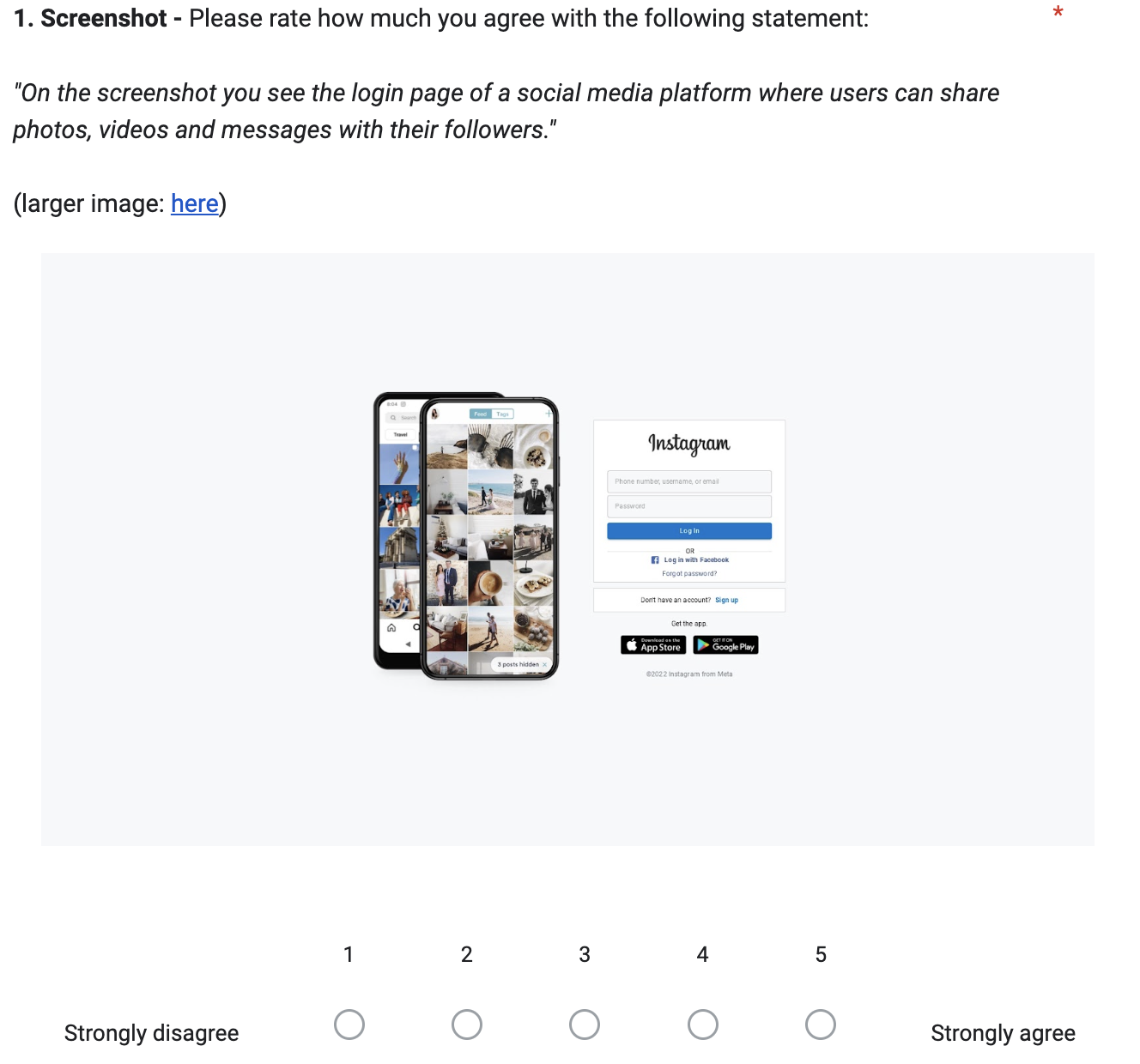}
    \caption{Exemplary question (i.e., the first) in part II of our questionnaire. The screenshot refers to an adversarial webpage.}
    \label{fig:question}
\end{figure}

\textbf{Design.} 
To reach our goals and allow answering our RQ, we created a semi-structured questionnaire~\cite{horton2004qualitative}, which enables both qualitative and quantitative analyses. Our questionnaire is divided into three parts ($\Rightarrow$ is the motivation):

\begin{enumerate}[label=\Roman*)]
    \item \textit{Demographics.} We ask preliminary questions, such as age, gender, and country of residence; but also education, expertise with IT and familiarity with some popular brands in Europe.
    $\Rightarrow$~We need this (non-PII~\cite{krishnamurthy2009leakage}) data to carry out fine-grained analyses, but also to comply with our third design goal, and with the European laws~\cite{fra2020child} (e.g., we manually deleted responses from people who were too young).
    
    \item \textit{Agreement.} We ask \smamath{20} closed questions having a similar format. Specifically, in each question we show the screenshot of a webpage, and then ask the participant to answer a question with the following phrasing: ``In the screenshot, it is shown a webpage of a popular brand. Do you agree with this statement?'', to which the user could respond in a [\smamath{1}--\smamath{5}] Linkert scale (with \smamath{1}=``Disagree'', and \smamath{5}=``Agree'').
    $\Rightarrow$~To comply with our second design goal, we try to avoid raising suspicion and ask a ``neutral'' question. Intuitively, if the participant agrees (i.e., answers with a \smamath{4} or a \smamath{5}), then it means that they believe the webpage to be genuine. We provide an exemplary question of part II in Fig.~\ref{fig:question}.

    \item \textit{Explanation.} We ask \smamath{20} open questions, containing the same webpages shown in the previous part. Specifically, we ask the users to explain ``why'' they disagreed (if so) with the statement written for the corresponding webpage.
    $\Rightarrow$~These questions are meant to investigate what made users ``suspicious'' of a given webpage. This is important to determine if there are any phishing elements that are noticeable by humans, but imperceptible to ML models. 
\end{enumerate}
The questionnaire ended with with a last, binary question, asking whether the participant ``changed their mind'' about some of the answers given in the second part. We did not impose any time limit to answer any of our questions (because, in reality, users do not have such a constraint), albeit we invited not to spend more than \smamath{15}s for questions in part II, and \smamath{1}m for those in part III (a similar ``soft-timer'' was used also in~\cite{lee2023attacking}).

\textbf{Content.} 
Our semi-structured questionnaire inquires the participants' opinion on 20 different webpages (in the form of screenshot) related to 12 brands popular in Europe. Of these 20 webpages, 2 are legitimate, which we included as a form of control (which we took by ourselves in early 2023); whereas the remaining 18 are AW. Importantly, the distribution of the screenshots in our questionnaire is \textit{fixed}: such a choice is to ensure consistency, but also to further avoid priming. Indeed, we put the AW that are more likely to raise suspicion (due to, e.g., clearly different logos) \textit{at the end of the questionnaire}: intuitively, if participants were shown suspicious webpages from the beginning, then they would be more skeptical of the remaining webpages---thereby leading to biased results. The 18 AW, as well as their placement in our questionnaire, were chosen after many meetings (some of which included \companyname's employees), in which the authors discussed the peculiar characteristics of each AW and eventually reached a consensus on a qualitative ``difficulty level'' to identify an AW as phishing. We report in Table~\ref{tab:questionnaire} the sequence of our screenshots, their brand and their difficulty level.

\textbf{Data collection.} 
We shared our questionnaire on popular social media (e.g., LinkedIn\footnote{We provided the questionnaire both in English and in German to enable even people with limited knowledge of English to participate.}). To avoid priming, we advertised it as ``Website Content Perception Survey''. We did not provide any payment to participants. We began collecting answers\footnote{We conducted \textbf{pilot} tests with colleagues. After submitting their responses, we revealed that 18 out of 20 screenshots were phishing. These pilot participants did not expect this, and some stated to be ``embarassed'' for being unable to figure this out.} on May 18th, 2023; and stopped after three weeks. We provide our complete questionnaire in our repository~\cite{repository}. Ethical remarks are discussed in §\ref{sec:conclusions}. 

\begin{figure*}[!htbp]
    \centering
    \begin{subfigure}[!htbp]{0.65\columnwidth}
        \centering
        \includegraphics[width=1\columnwidth]{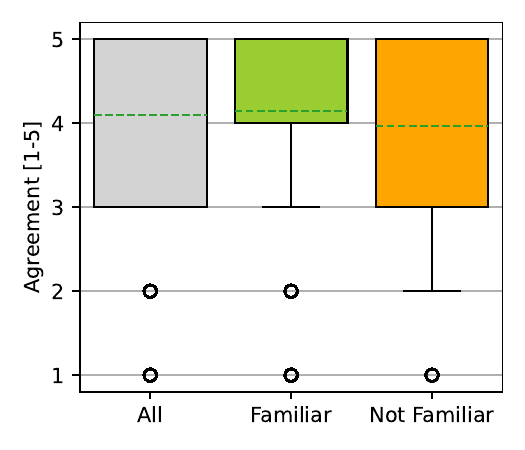}
        \caption{Aggregated ratings (only AW).}
        \label{sfig:aggregated_overall}
    \end{subfigure}
    \begin{subfigure}[!htbp]{0.65\columnwidth}
        \centering
        \includegraphics[width=1\columnwidth]{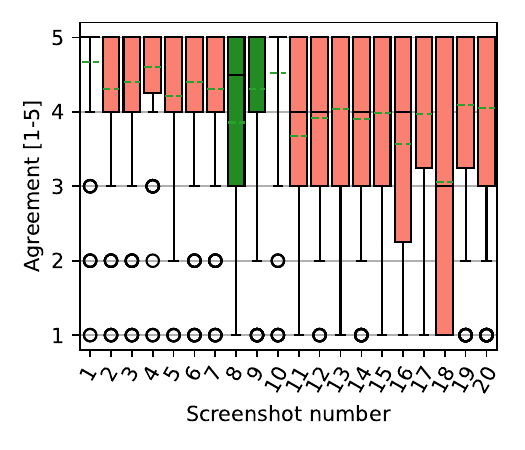}
        \caption{Rating per screenshot (entire sample).}
        \label{sfig:individual_all}
    \end{subfigure}
    \begin{subfigure}[!htbp]{0.65\columnwidth}
        \centering
        \includegraphics[width=1\columnwidth]{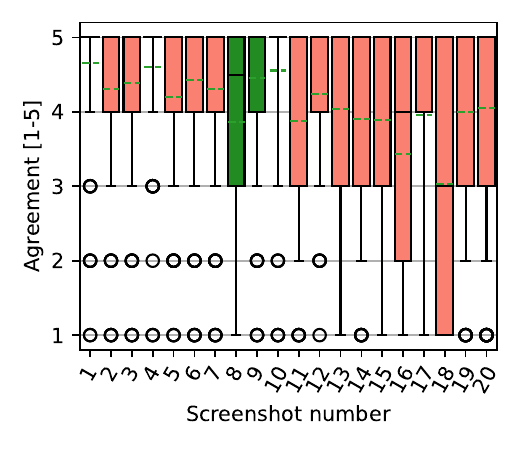}
        \caption{Rating per screenshot (only familiar with brand).}
        \label{sfig:individual_familiar}
    \end{subfigure}
    \caption{High-level results. Fig.~\ref{sfig:aggregated_overall} reports the aggregated distribution for the 18 AW in our questionnaire. Figs.~\ref{sfig:individual_all} and~\ref{sfig:individual_familiar} show the rating distribution for each screenshot (green for legitimate, red for AW). Our AW can deceive most users (especially at the start of the questionnaire).}
    \label{fig:main_results}
\end{figure*}

\subsection{Sample and Limitations}
\label{ssec:sample}
\noindent
We describe our sample and the limitations of our research.

\textbf{Sample Description.}
We received 126 responses. \textit{Gender}-wise, 70 (\smamath{55.6\%}) identified themselves as male, and 55 (\smamath{43.6\%}) as female (1 did not answer). In terms of \textit{age}, 3 (\smamath{2.4\%}) are younger than 16; 44 (\smamath{34.9\%}) are in the 16--24 range;
57 (\smamath{45.2\%}) between 25--34;  12 (\smamath{9.5\%}) between 35--44; 4 (\smamath{3.1\%}) between 45--54; 6 (\smamath{4.8\%}) between 55--64; and none are older than 65. With regards to \textit{country of residence}, 70 (\smamath{61.9\%}) are from Austria; 19 (\smamath{15.1\%}) are from Germany, and 12 (\smamath{9.5\%}) from Switzerland; 5 (\smamath{3.9\%}) are from Bosnia, 4 (\smamath{3.2\%}) are from Slovenia, and 2 (\smamath{1.6\%}) from Liechtenstein; 1 (0.8\%) are from Finland, Georgia, Macedonia, Estonia, Italy, as well as from the USA. The \textit{educational} background of the participants accentuated the diverse nature of our sample: 45 (\smamath{35.7\%}) have a high-school diploma, whereas 41 (\smamath{32.5\%}) a BSc, and 27 (\smamath{21.4\%}) an MSc; 2 (\smamath{1.6\%}) have a PhD; 11 (\smamath{8.7\%}) only completed basic schooling. With respect to \textit{expertise with IT}, 75 (\smamath{59.5\%}) participants are heavily involved with IT (either professionally or for personal interests), whereas 48 (\smamath{38.1\%}) only use IT for entertainment or when necessary; lastly, 3 (\smamath{2.4\%}) reported to make a very limited use of IT in their daily lives. Finally, we report in Table~\ref{tab:familiarity} the amount of participants that are familiar with the brands we considered, showing that most of our sample is familiar with our chosen brands---validating the real-world applicability of our findings.

\begin{table}[!htbp]
\caption{Familiarity of our sample with the twelve brands in our questionnaire. On average, our brands are known by 91 (72\%) participants.}
\label{tab:familiarity}

\resizebox{0.95\columnwidth}{!}{

\begin{tabular}{c|cccccccccccc}
\toprule
 \textbf{Brand}
 & \rotatebox[origin=c]{90}{NetFlix} 
 & \rotatebox[origin=c]{90}{Amazon} 
 & \rotatebox[origin=c]{90}{Zalando} 
 & \rotatebox[origin=c]{90}{AirBnB} 
 & \rotatebox[origin=c]{90}{Google} 
 & \rotatebox[origin=c]{90}{Instagram} 
 & \rotatebox[origin=c]{90}{Facebook} 
 & \rotatebox[origin=c]{90}{LinkedIn} 
 & \rotatebox[origin=c]{90}{PayPal} 
 & \rotatebox[origin=c]{90}{Uber} 
 & \rotatebox[origin=c]{90}{Yahoo} 
 & \rotatebox[origin=c]{90}{Twitter} \\
\midrule
\textbf{Absolute} & \ftmath{117} & \ftmath{109} & \ftmath{92} & \ftmath{80} & \ftmath{123} & \ftmath{117} & \ftmath{98} & \ftmath{96} & \ftmath{93} & \ftmath{55} & \ftmath{42} & \ftmath{68} \\
\textbf{Relative} & \ftmath{93\%} & \ftmath{87\%} & \ftmath{73\%} & \ftmath{63\%} & \ftmath{98\%} & \ftmath{93\%} & \ftmath{78\%} & \ftmath{76\%} & \ftmath{74\%} & \ftmath{44\%} & \ftmath{33\%} & \ftmath{54\%} \\

\bottomrule
\end{tabular}
}
\end{table}

\textbf{Limitations.}
Our sample and questionnaire have some intrinsic limitations. For instance, most of our participants are from German-speaking countries, so our study is biased towards this area. Furthermore, our sample exhibits significant diversity in terms of age, gender and background: this characteristic is both a strength (since it allows deriving broad takeaways) and a weakness (since it impairs specificity). Finally, our questionnaire includes only 18 AW pertaining to some reputable brands: therefore, there may be other brands, or other types of AW, for which our study cannot provide any answer. For these reasons, \textit{we do not claim that the results of our research can be generalized} to reflect the entire human population\footnote{We also observe that our sample is larger than the one considered by most prior work (c.f. Table~\ref{tab:related}), and that even recent top-conferences accepted papers (e.g.,~\cite{alahmadi202299}) having user-studies with a smaller population than ours.} and/or the full landscape of phishing threats. Nonetheless, given the exploratory nature of our study (which is the first to investigate our RQ) as well as the many precautions we took to reduce priming and ensure realistic assessments, our results represent a significant step towards mitigating the proliferation of phishing websites.

\begin{figure*}[!htbp]
    \centering
    \begin{subfigure}[!htbp]{0.49\columnwidth}
        \centering
        \includegraphics[width=1\columnwidth]{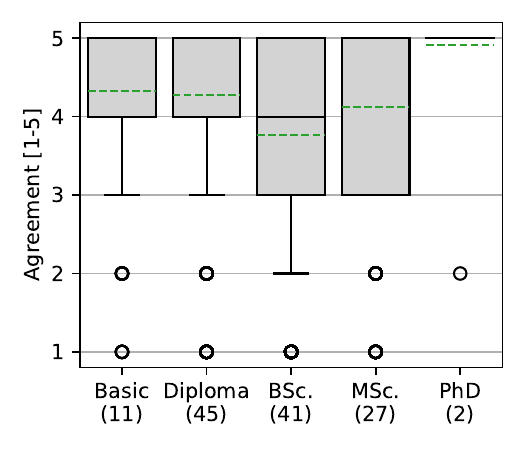}
        \caption{Education.}
        \label{sfig:aggregated_education}
    \end{subfigure}
    \begin{subfigure}[!htbp]{0.49\columnwidth}
        \centering
        \includegraphics[width=1\columnwidth]{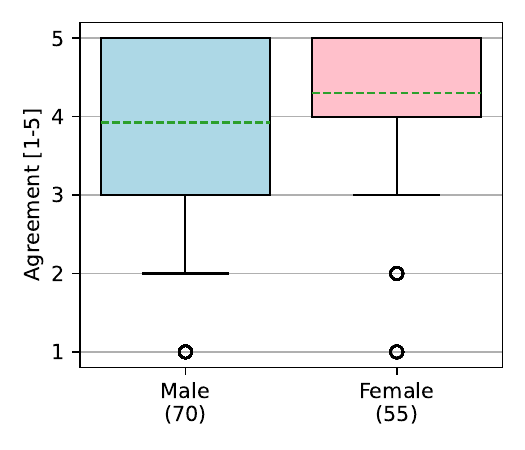}
        \caption{Gender.}
        \label{sfig:aggregated_gender}
    \end{subfigure}
    \begin{subfigure}[!htbp]{0.49\columnwidth}
        \centering
        \includegraphics[width=1\columnwidth]{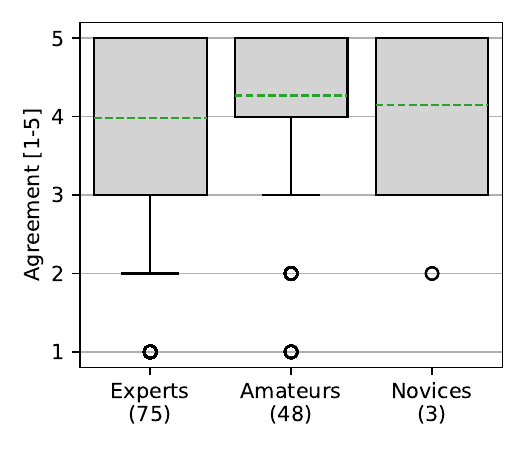}
        \caption{Expertise with IT.}
        \label{sfig:aggregated_expertise}
    \end{subfigure}
    \begin{subfigure}[!htbp]{0.49\columnwidth}
        \centering
        \includegraphics[width=1\columnwidth]{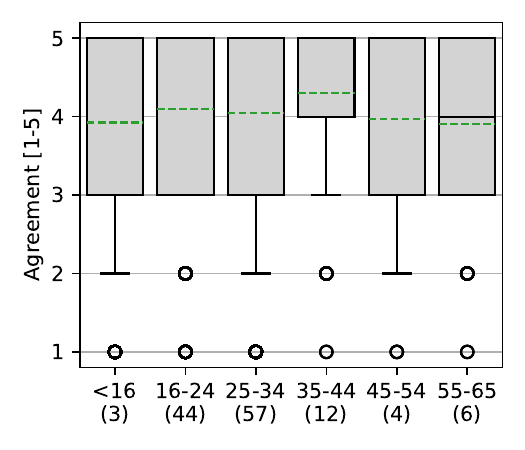}
        \caption{Age.}
        \label{sfig:aggregated_age}
    \end{subfigure}
    \caption{Subgroup results. The figures report the aggregated ratings (for the 18 AW) of each subgroup (the x-axis denotes the size of each subgroup).}
    \label{fig:subgroups_results}
\end{figure*}
\section{Results (quantitative)}
\label{sec:results}
\noindent
We now report the quantitative results of our user-study (from part~II). We first show high-level findings (§\ref{ssec:general}), and then focus on specific subsets of our sample (§\ref{ssec:group}). We also analyze two intriguing phenomena on the natural progression of our questionnaire (§\ref{ssec:progression}). Finally, we perform fine-grained analyses on some relevant combinations of our sample's demographics (§\ref{ssec:finegrained}). While presenting the results, we will make some ``claims'' (denoted as [\smamath{\mathbb{C}}]), which we validate through statistical tests in the next section (§\ref{ssec:statistical}).

\subsection{General findings}
\label{ssec:general}
\noindent
We begin by addressing our main RQ at a high-level. We report in Fig.~\ref{sfig:aggregated_overall} the distribution of the ``agreement ratings'' that our participants provided for \textit{all} the 18 AW screenshots in our questionnaire. (As a reminder, \textbf{higher agreement implies higher likelihood of being deceived}.) Specifically, we provide three boxplots: the leftmost one represents our entire sample (having 2268 ratings, given by 126 participants~*~18 AW); the central one represents the ratings provided only by those participants who reported being ``familiar'' with the corresponding brand of each AW (having 1666 ratings); whereas the rightmost one represents the ratings of those participants who stated otherwise (having 602 ratings). From Fig.~\ref{sfig:aggregated_overall}, we can see that the ratings of most of our sample are above the saddle point (of 3), and the mean for these three boxplots is \smamath{\sim}4. These results suggest that our chosen AW \textit{can deceive the human users}. Surprisingly, it appears that users who are \textit{more familiar} with a given brand tend to be \textit{easier to deceive} [\smamath{\mathbb{C}1}].

To provide a deeper understanding, we report the ratings provided to each individual screenshot (AW are in red, whereas legitimate websites are in green) by our entire sample (Fig.~\ref{sfig:individual_all}) and by only those who are ``familiar'' with the brand of the screenshot (Fig.~\ref{sfig:individual_familiar}). We see that ``familiar'' participants are worse at identifying the phishing nature of screenshots \#4, \#13 and \#17. Finally, we see that the rating for the first five AW is very high (around \smamath{4.5} on average) meaning that \textbf{these AW can successfully ``phish''} most of our sample---under the assumption that {\small \textit{(i)}} no additional contextual information is provided and that {\small \textit{(ii)}} there is no priming.

\begin{cooltextbox}
\textsc{\textbf{Takeaway.}}
Most of our sample cannot recognize AW, and familiarity with a brand hinders the detection skills of users.
\end{cooltextbox}

\subsection{Group-specific analyses}
\label{ssec:group}
\noindent
We now focus our attention on specific subsets of our sample. We rely on the demographics information provided by participants at the beginning. We do so by providing the aggregated rating distribution (only for the 18 AW) of our sample on the basis of: education (Fig.~\ref{sfig:aggregated_education}), gender (Fig.~\ref{sfig:aggregated_gender}), expertise with IT (Fig.~\ref{sfig:aggregated_expertise}), and age (Fig.~\ref{sfig:aggregated_age}); the x-axes of all figures in Fig.~\ref{fig:subgroups_results} report the population of each subgroup.

By observing Figs.~\ref{fig:subgroups_results}, we can make four significant \textbf{claims}:
\begin{enumerate}[label={[}\smamath{\mathbb{C}}\arabic*{]}]
\setcounter{enumi}{1}
    \item University graduates are more suspicious (Fig.~\ref{sfig:aggregated_education}).
    \item Female appear to be less suspicious than males (Fig.~\ref{sfig:aggregated_gender}).
    \item IT experts are more skeptical than amateurs (Fig.~\ref{sfig:aggregated_expertise}).
    \item Age is not correlated with suspiciousness (Fig.~\ref{sfig:aggregated_age}).
\end{enumerate}
Regardless, we found it intriguing that those who possess a PhD tend to be the easiest to be deceived---albeit we cannot support such a claim, since only two participants of our sample have a doctorate.

\subsection{Skepticism over-time}
\label{ssec:progression}

\begin{figure}[!htbp]
    \centering
    \begin{subfigure}[!htbp]{0.49\columnwidth}
        \centering
        \includegraphics[width=1\columnwidth]{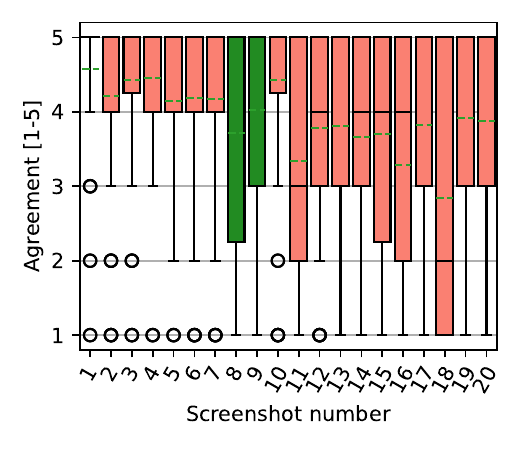}
        \caption{Male (N=70).}
        \label{sfig:individual_male}
    \end{subfigure}
    \begin{subfigure}[!htbp]{0.49\columnwidth}
        \centering
        \includegraphics[width=1\columnwidth]{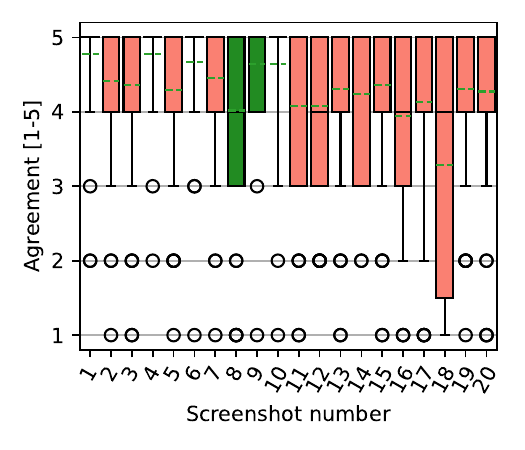}
        \caption{Female (N=55).}
        \label{sfig:individual_female}
    \end{subfigure}
    \caption{Individual screenshot ratings based on Gender.}
    \label{fig:male_vs_female}
\end{figure}

\begin{figure}[!htbp]
    \centering
    \begin{subfigure}[!htbp]{0.49\columnwidth}
        \centering
        \includegraphics[width=1\columnwidth]{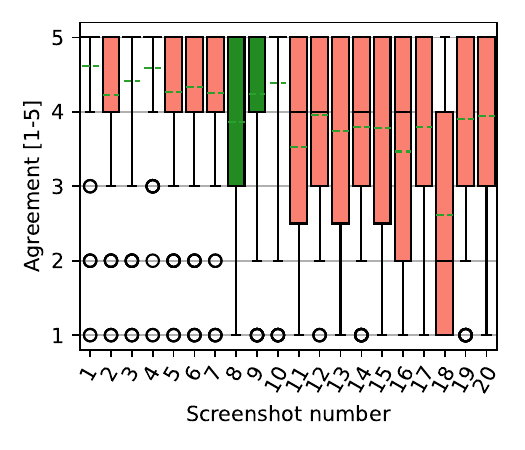}
        \caption{IT experts (N=75).}
        \label{sfig:individual_expert}
    \end{subfigure}
    \begin{subfigure}[!htbp]{0.49\columnwidth}
        \centering
        \includegraphics[width=1\columnwidth]{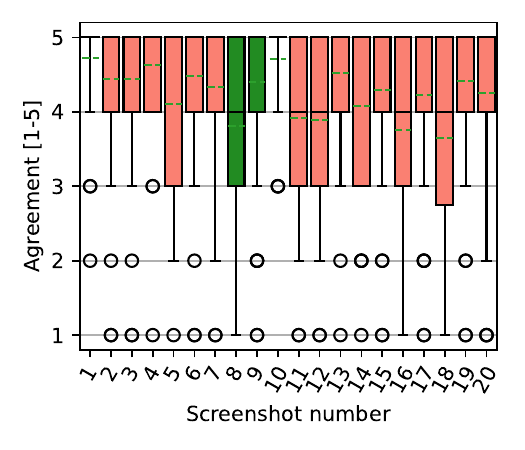}
        \caption{IT amateurs (N=48).}
        \label{sfig:individual_amateur}
    \end{subfigure}
    \caption{Individual screenshot ratings based on Expertise with IT.}
    \label{fig:passionate_vs_amateurs}
\end{figure}

\noindent
We find it instructive to analyze the natural progression of the ratings as each participant advanced in the questionnaire. Indeed, we recall (§\ref{ssec:questionnaire}) that -- to avoid priming -- we placed the hardest AW to identify at the beginning of the questionnaire, whereas the easiest were at the end. Hence, our participants are bound to become more suspicious over-time, which would lead to a drop in the agreement ratings. 
We perform this exercise by focusing on two demographics: gender (Figs.~\ref{fig:male_vs_female}) and expertise with IT (Figs.~\ref{fig:passionate_vs_amateurs}).

\textbf{Male vs Females.} By comparing Fig.~\ref{sfig:individual_male} with Fig.~\ref{sfig:individual_female}, we can see that both groups exhibit high agreement (avg\smamath{\approx\!4.3} for males, and \smamath{\approx\!4.5} for females) in the first 7 screenshots---all being AW. (Interestingly, males tend to be dubious of the two legitimate screenshots!) However, starting from 10th screenshot, the agreement of males starts decreasing (avg\smamath{\approx\!3.7}), whereas those of females remains high (avg\smamath{\approx\!4.2}) until screenshot \#18, which leads to a drop in agreement by most (avg\smamath{\approx\!2.8} for males, and \smamath{\approx\!3.3} for females).

\textbf{Experts vs Amateurs.}\footnote{We use ``amateur'' to denote participants who ``use IT only when necessary or for entertainment'', and ``expert'' for those who are ``passionate about IT''.} By comparing Fig.~\ref{sfig:individual_expert} with Fig.~\ref{sfig:individual_amateur}, we see some interesting trends. Specifically, ``experts'' tend to agree similarly to ``amateurs'' at the beginning [\smamath{\mathbb{C}6}]; however, after completing half of the questionnaire, ``experts'' become much more skeptical than ``amateurs'' [\smamath{\mathbb{C}7}].

\begin{figure*}[!htbp]
    \centering
    \begin{subfigure}[!htbp]{0.49\columnwidth}
        \centering
        \includegraphics[width=1\columnwidth]{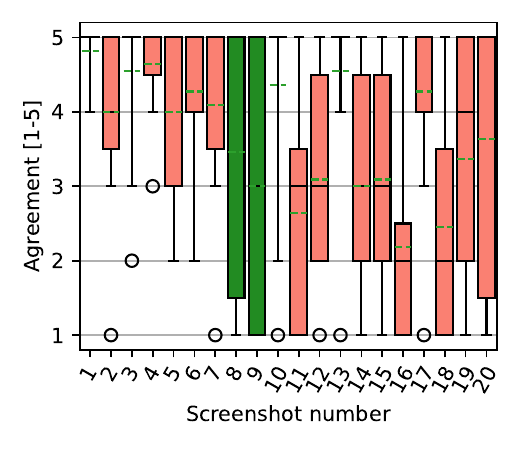}
        \caption{Male, IT Experts (N=11).}
        \label{sfig:individual_male_1624_expert}
    \end{subfigure}
    \begin{subfigure}[!htbp]{0.49\columnwidth}
        \centering
        \includegraphics[width=1\columnwidth]{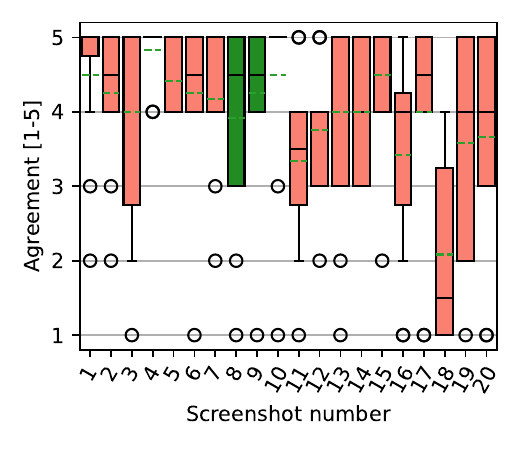}
        \caption{Female, IT Experts (N=12).}
        \label{sfig:individual_female_1624_expert}
    \end{subfigure}
    \begin{subfigure}[!htbp]{0.49\columnwidth}
        \centering
        \includegraphics[width=1\columnwidth]{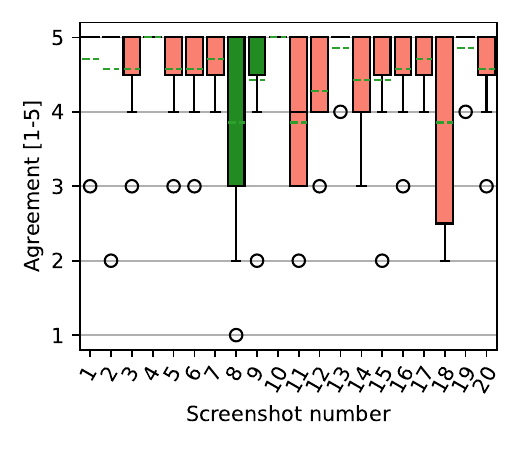}
        \caption{Male, IT amateurs (N=7).}
        \label{sfig:individual_male_1624_amateur}
    \end{subfigure}
    \begin{subfigure}[!htbp]{0.49\columnwidth}
        \centering
        \includegraphics[width=1\columnwidth]{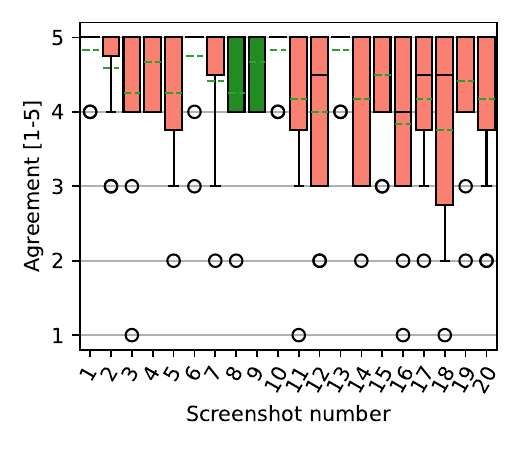}
        \caption{Female, IT amateurs (N=12).}
        \label{sfig:individual_female_1624_amateur}
    \end{subfigure}
    \caption{Case-study. Individual screenshot ratings of participants aged 16--24 (N=44), categorized on the basis of gender and IT expertise.}
    \label{fig:casestudy_1624}
\end{figure*}

\begin{figure*}[!htbp]
    \centering
    \begin{subfigure}[!htbp]{0.49\columnwidth}
        \centering
        \includegraphics[width=1\columnwidth]{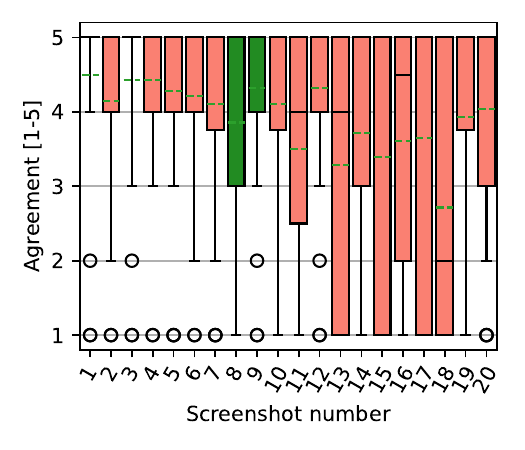}
        \caption{Male, IT Experts (N=28).}
        \label{sfig:individual_male_2534_expert}
    \end{subfigure}
    \begin{subfigure}[!htbp]{0.49\columnwidth}
        \centering
        \includegraphics[width=1\columnwidth]{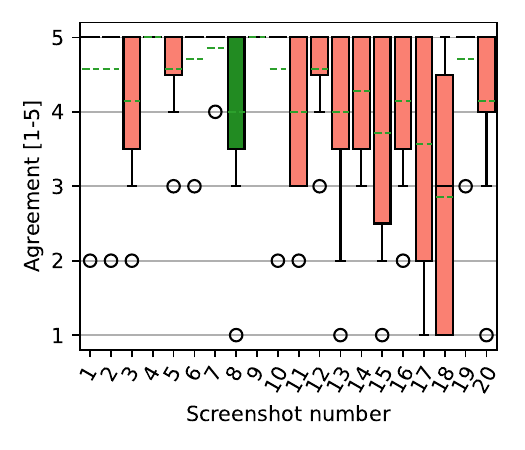}
        \caption{Female, IT Experts (N=7).}
        \label{sfig:individual_female_2534_expert}
    \end{subfigure}
    \begin{subfigure}[!htbp]{0.49\columnwidth}
        \centering
        \includegraphics[width=1\columnwidth]{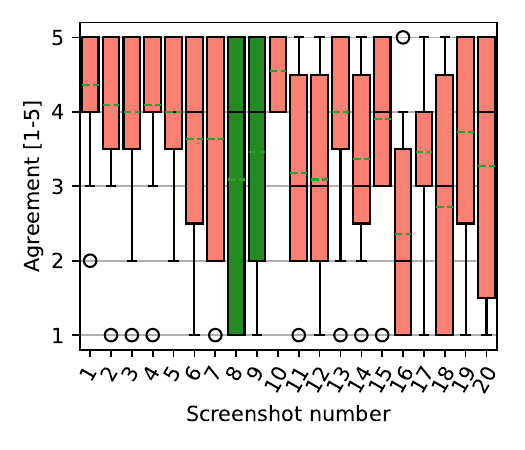}
        \caption{Male, IT amateurs (N=11).}
        \label{sfig:individual_male_2534_amateur}
    \end{subfigure}
    \begin{subfigure}[!htbp]{0.49\columnwidth}
        \centering
        \includegraphics[width=1\columnwidth]{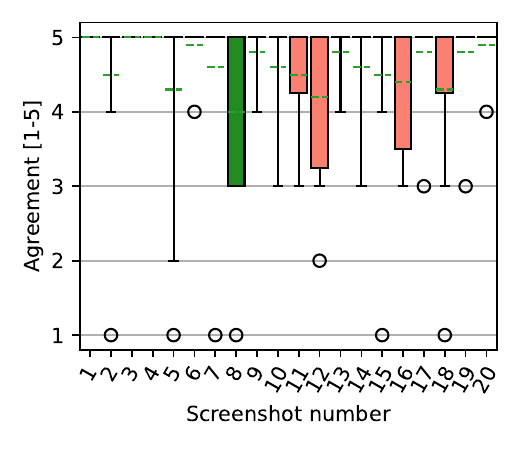}
        \caption{Female, IT amateurs (N=10).}
        \label{sfig:individual_female_2534_amateur}
    \end{subfigure}
    \caption{Case-study. Individual screenshot ratings of participants aged 25--34 (N=57), categorized on the basis of gender and IT expertise.}
    \label{fig:casestudy_2534}
\end{figure*}

\begin{cooltextbox}
\textsc{\textbf{Takeaway.}}
As participants advance in our questionnaire, they appear to become more suspicious.
\end{cooltextbox}

\subsection{Fine-grained analyses}
\label{ssec:finegrained}
\noindent
We conclude our results by focusing our attention at various subgroups of our sample, which we draw from those having the highest amount of participants. Specifically, we consider the individual ratings of participants who are aged 16--24 (Figs.~\ref{fig:casestudy_1624}) or 25--34 (Figs.~\ref{fig:casestudy_2534}) and that are either male or female, and either IT experts or amateurs---thereby resulting in eight different combinations. While we acknowledge that some of them have few examples, we find it educational to analyze these case-studies---which can be considered extensions of those discussed in §\ref{ssec:progression},

By observing Figs.~\ref{fig:casestudy_1624} and Figs.~\ref{fig:casestudy_2534}, we derive the following:
\begin{itemize}
    \item Female IT amateurs become less skeptical as they age (cf. Fig.~\ref{sfig:individual_female_1624_amateur} with ~\ref{sfig:individual_female_2534_amateur})...
    \item ...but the opposite holds for males (cf. Fig.~\ref{sfig:individual_male_1624_amateur} with~\ref{sfig:individual_male_2534_amateur}).
    \item Legitimate webpages appear suspicious to male IT amateurs aged 25--34, but not for 16--24 (cf. Fig.~\ref{sfig:individual_male_2534_amateur} with~\ref{sfig:individual_male_1624_amateur})...
    \item ...but the opposite holds for male IT experts: those aged 16--24 are more suspicious of legitimate webpages w.r.t. those aged 25--34 (cf. Fig.~\ref{sfig:individual_male_1624_expert} with Fig.~\ref{sfig:individual_male_2534_expert}).
    \item Female IT amateurs aged 25--34 have the highest ``agreement'' ratings---making them more susceptible to AW.
\end{itemize}
Given the small sample size, we refrain from making claims on the these observations. However, the \textbf{lesson learnt} is that some groups of users are more vulnerable to AW than others.

\section{Validation and Analysis}
\label{sec:analysis}
\noindent
We expand our quantitative analyses on part~II.
We validate our claims (§\ref{ssec:statistical}) and draw similarities with prior work (§\ref{ssec:comparison}).

\subsection{Statistical validation}
\label{ssec:statistical}
\noindent
We validate our 7 claims made in §\ref{sec:results} through statistical tests. 

\textbf{Hypotheses.} Inspired by prior work~\cite{baki2022sixteen}, we rely on the Welch's t-test~\cite{delacre2017psychologists}. This test can determine if two groups are equal by comparing the resulting \smamath{p}-value with a given target \smamath{\alpha} (typically set to \smamath{0.05}). Hence, for each claim, we identify two groups (\smamath{g1} and \smamath{g2}), compute the \smamath{p}-value, and use it to test the null hypothesis (\smamath{H_0}): ``\smamath{g1} and \smamath{g2} are statistically equivalent''. \smamath{H_0} is accepted if \smamath{p\!>\!0.05}, and rejected otherwise. It can also be that the test is inconclusive (due to, e.g., lack of data-points): to provide more confident conclusions, we also measure the effect size (ES) of each test.

\textbf{Setup.} All our claims refer to how our participants analyzed phishing screenshots. For each test, our groups entail the agreement ratings provided by the two compared groups (\smamath{g1} and \smamath{g2}) to a specified set of AW screenshots. Let us identify the groups we considered in our tests to validate each claim.

\begin{enumerate}[label=\smamath{\mathbb{C}}\arabic*:,leftmargin=0.6cm]
    \item \textit{Familiarity.} (\smamath{H_0} should be rejected); \smamath{g1} denotes participants who are familiar with the brand of a given AW, whereas \smamath{g2} denotes those who are not familiar.
    
    \item \textit{University.} (\smamath{H_0} should be rejected) \smamath{g1} denotes participants with a degree (BSc., MSc., PhD), whereas \smamath{g2} are those without a degree (Basic school or high-school).

    \item \textit{Gender.} (\smamath{H_0} should be rejected) \smamath{g1} denotes those who identified as male, and \smamath{g2} as female.

    \item \textit{IT expertise.} (\smamath{H_0} should be rejected) \smamath{g1} denotes those who are ``experts'', and \smamath{g2} ``amateurs''. 

    \item \textit{Age.} (\smamath{H_0} should be accepted) \smamath{g1} denotes participants aged \smamath{<}25 (47 in total), \smamath{g2} those aged 25--34 (57); we also consider \smamath{g3} including those \smamath{>}34 (22).

    \item \textit{Similar beginning.} (\smamath{H_0} should be accepted) We consider the first 4 AW;  \smamath{g1} denotes experts in IT, and \smamath{g2} amateurs. 

    \item \textit{Different ending.} (\smamath{H_0} should be rejected.) We consider the last 10 AW; \smamath{g1} denotes experts in IT, and \smamath{g2} amateurs. 
    
\end{enumerate}
\noindent 
For [\smamath{\mathbb{C}1}]--[\smamath{\mathbb{C}5}], \smamath{g1} and \smamath{g2} include all 18 AW.

\textbf{Results.} We display the results of these tests in Table~\ref{tab:statistical}, in which rows report the amount of elements (N), the average and standard deviation of each group; as well as the \smamath{p}-value (green/red cells denote cases in which \smamath{H_0} must be accepted/rejected) and the ES of the test.
Table~\ref{tab:statistical} shows that \textit{all our claims are validated}: cases in which \smamath{H_0} must be rejected also show a small ES, which provides additional evidence that the two groups are statistically different. Finally, for \smamath{\mathbb{C}5} (for which we identified 3 groups), we also compared \smamath{g2} with \smamath{g3} (having avg=\smamath{4.07}, std=\smamath{1.13}, N=\smamath{396}) and we find that \smamath{p}=\smamath{0.42} (ES=\smamath{0.047}) i.e., \smamath{H_0} must be accepted (since \smamath{p\!>\!\alpha}=\smamath{0.05}); hence, since \smamath{g1}\smamath{\equiv}\smamath{g2}, and \smamath{g2}\smamath{\equiv}\smamath{g3}, it follows that \smamath{g1}\smamath{\equiv}\smamath{g3}, which validates the claim that age has a negligible impact on phishing awareness---at least according to our sample.

\begin{table}[!htbp]
\caption{Statistical validation of our claimed hypotheses ($\alpha$=$0.05$)}
\label{tab:statistical}

\resizebox{1\columnwidth}{!}{

\begin{tabular}{c|cc|cc|cc|cc|cc|cc|cc}
\toprule

\multirow{2}{*}{\textbf{Claim}} & \multicolumn{2}{c|}{\smamath{\mathbb{C}1}} & \multicolumn{2}{c|}{\smamath{\mathbb{C}2}} & \multicolumn{2}{c|}{\smamath{\mathbb{C}3}} & \multicolumn{2}{c|}{\smamath{\mathbb{C}4}} & \multicolumn{2}{c|}{\smamath{\mathbb{C}5}} & \multicolumn{2}{c|}{\smamath{\mathbb{C}6}} & \multicolumn{2}{c}{\smamath{\mathbb{C}7}} \\ \cline{2-15}
& \smamath{g1} & \smamath{g2} & \smamath{g1} & \smamath{g2} & \smamath{g1} & \smamath{g2} & \smamath{g1} & \smamath{g2} & \smamath{g1} & \smamath{g2} & \smamath{g1} & \smamath{g2} & \smamath{g1} & \smamath{g2} \\
\midrule
N & \ftmath{1666} & \ftmath{602} & \ftmath{1260} & \ftmath{1008} & \ftmath{1260} & \ftmath{990} & \ftmath{1350} & \ftmath{864} & \ftmath{846} & \ftmath{1080} & \ftmath{300} & \ftmath{192} & \ftmath{750} & \ftmath{480} \\

avg. & \ftmath{4.14} & \ftmath{3.96} & \ftmath{3.94} & \ftmath{4.29} & \ftmath{3.92} & \ftmath{4.30} & \ftmath{3.98} & \ftmath{4.27} & \ftmath{4.13} & \ftmath{4.07} & \ftmath{4.46} & \ftmath{4.56} & \ftmath{3.65} & \ftmath{4.10} \\

std. & \ftmath{1.31} & \ftmath{1.29} & \ftmath{1.38} & \ftmath{1.18} & \ftmath{1.43} & \ftmath{1.10} & \ftmath{1.41} & \ftmath{1.13} & \ftmath{1.22} & \ftmath{1.41} & \ftmath{1.03} & \ftmath{0.88} & \ftmath{1.54} & \ftmath{1.23} \\ \hline

$p$ & \multicolumn{2}{c|}{\cellcolor{red!10}\smamath{0.004}} & \multicolumn{2}{c|}{\cellcolor{red!10}\smamath{<0.001}} & \multicolumn{2}{c|}{\cellcolor{red!10}\smamath{<0.001}} & \multicolumn{2}{c|}{\cellcolor{red!10}\smamath{<0.001}} & \multicolumn{2}{c|}{\cellcolor{green!10}\smamath{0.32}} & \multicolumn{2}{c|}{\cellcolor{green!10}\smamath{0.26}} & \multicolumn{2}{c}{\cellcolor{red!10}\smamath{<0.001}}\\

ES & \multicolumn{2}{c|}{\smamath{0.14}} & \multicolumn{2}{c|}{\smamath{0.27}} & \multicolumn{2}{c|}{\smamath{0.29}} & \multicolumn{2}{c|}{\smamath{0.22}} & \multicolumn{2}{c|}{\smamath{0.046}} & \multicolumn{2}{c|}{\smamath{0.1}} & \multicolumn{2}{c}{\smamath{0.31}} \\

 \bottomrule
\end{tabular}
}

\end{table}

\textbox{\textbf{Remark:} All our claims and findings pertain to the data of our user-study. We do not generalize (see §\ref{ssec:sample}).}

\subsection{Comparison with prior work}
\label{ssec:comparison}
\noindent
We appreciate that our sample bears some resemblance with the one of prior studies~\cite{baki2022sixteen}. For instance, our \textit{gender} distribution is similar to the user-studies in~\cite{lastdrager2017effective,orunsolu2018users,alsharnouby2015phishing}. Interestingly, Orunsolu et al.~\cite{orunsolu2018users} found that female perform better (their sample had a male:female split of 57:43), while the study by Kumaraguru et al.~\cite{kumaraguru2010teaching} found otherwise (albeit the 28 participants in~\cite{kumaraguru2010teaching} had a male-to-female ratio of 5-to-1). Despite having a different goal in mind, our findings align with those in~\cite{kumaraguru2010teaching} (see \smamath{\mathbb{C}3}).

From an \textit{age} perspective, Purkait et al.~\cite{purkait2014empirical} (whose sample was spread between 20--62 years of age) found that elders were more susceptible; the opposite was found by Kumaraguru~\cite{kumaraguru2010teaching} (whose sample was aged 13--65), who claimed that participants between 13--17 were the most susceptible to phishing---a finding shared also by Lastdrager et al.~\cite{lastdrager2017effective}. In contrast, we did not find any significant performance difference related to age (see \smamath{\mathbb{C}5})---a result that aligns with those in~\cite{sheng2007anti, herzberg2008security, gopavaram2021cross}.

Finally, the user-studies in~\cite{alsharnouby2015phishing, sheng2007anti} found that \textit{expertise with IT} was not correlated with phishing susceptibility. In contrast, Orunsulu et al.~\cite{orunsolu2018users} found that experts are more resilient---which aligns with our \smamath{\mathbb{C}4} (albeit this may not hold for specific subgroups §\ref{ssec:finegrained}). We stress, however, that drawing conclusions based on similar correlations may be misleading---as echoed in a very recent work~\cite{lopez2022role}.

\vspace{1mm}

\textbox{\textbf{Remark:} our user-study has different goals than those by prior work (§\ref{ssec:related}). Hence, comparisons may not be appropriate, and we do these solely for educational purposes.}

\section{Explanations (qualitative)}
\label{sec:explanations}
\noindent
We qualitatively analyze the responses we received for part~III.

\subsection{Considered screenshots}
\label{ssec:considered}
\noindent
Our participants provided plenty of (unstructured) comments in part~III, and objectively analyzing all of these is impossible---given that such responses are also in diverse languages (which would further add bias in the translation). Hence, we prefer to focus on the responses we received for three meaningful screenshots, namely:
\begin{itemize}
    \item \textbf{Screenshot \#1} (Instagram---hard difficulty, shown in Fig.~\ref{fig:question}) $\Rightarrow$ This is the \textit{first} screenshot of our questionnaire. Hence, it represents the perfect use-case since there is no form of ``phishing priming'' that may influence the agreement of our participants. Perhaps unsurprisingly, the average rating for this screenshot is \smamath{\sim\!4.8} (cf. Fig.~\ref{sfig:individual_all}). 

    \item \textbf{Screenshot \#10} (Netflix---moderate difficulty, cf. Fig.~\ref{sfig:screenshot10}). $\Rightarrow$ This screenshot is at the middle of our questionnaire, and represents a good balance between difficulty (it is easier to identify than \#1) and priming (participants may have begun to become suspicious after answering the previous nine questions). Its average rating is \smamath{\sim\!\!4.5} (cf. Fig.~\ref{sfig:individual_all}).

    \item \textbf{Screenshot \#18} (Netflix---very easy difficulty, cf. Fig.~\ref{sfig:screenshot18}). $\Rightarrow$ This screenshot is at the end of our questionnaire, and aside from being very easy to identify as phishing, it also refers to Netflix (same as \#10). Analyzing \#18 is useful to investigate what elements of an AW are ``easily spotted'' by humans, thereby allowing practitioners to either ignore similar AW (since humans can easily recognize them as phishing) or focus on them (to avoid annoying users) to improve their PDS. Its average rating is \smamath{\sim\!3} (cf. Fig.~\ref{sfig:individual_all}).
\end{itemize}
To avoid bias due to translation, we base these analyses on the responses of \textbf{German-speaking participants}, which represent \smamath{82\%} (i.e., 103 out of 126) of our sample (see §\ref{ssec:sample}). 

\subsection{Assessment}
\label{ssec:words}
\noindent
Overall, 8 (\smamath{8\%}) rated screenshot \#1 with a 3 or less, and 18 (\smamath{17\%}) wrote a comment on the corresponding question in part~III; for screenshot \#10, 16 (\smamath{16\%}) rated it with a 3 or less, and 21 (\smamath{20\%}) wrote a comment on it; for screenshot \#18, 61 (\smamath{59\%}) rated it with a 3 or less, and 69 (\smamath{67\%}) wrote a comment.

Let us report some remarks written by our participants.
\begin{enumerate}[label=\#\arabic*:]
    \item the few comments ``confirm'' that participants agree with the statement. Others reported ``not having enough knowledge about Instagram to confirm certain statements''. The only valuable remarks from a phishing perspective are those by users who reported that the presentation of the webpage is ``incorrect'', and that there is a ``lack of the logo'' (which we found to be odd, since screenshot \#1 has the Instagram logo, and even the real login webpage of Instagram does not have another logo in it).

\setcounter{enumi}{9}
    \item many participants expressed concerns on the logo, which is described as being ``distorted'' or ``inauthentic''. Some mentioned weird placement of ``tabs'', and that the screenshot lacks a ``search function''. Few mentioned that NetFlix does not offer only ``movies'', but also ``documentaries'' (not shown in the screenshot). The ``font type'' and ``headings'' were also mentioned as source of doubt.

\setcounter{enumi}{17}
    \item many commented that the logo is ``incorrect'' or ``outdated''. Concerns were made on the overall look of the webpage, which appears ``cheap'', ``unprofessional'' and ``untrustworthy''. Specifically, some stated that it ``does not say anything about Netflix'' (i.e., there are no ``images or movies'') and that it only resembles a ``registration page''. The lack of a ``search function'' was also reported frequently. Some participants also criticized the ``color scheme'', which does not match the one of Netflix. 
\end{enumerate}
Given the abundant feedback we received for screenshot \#10 and \#18, we visualize these comments (in Fig.~\ref{fig:wordclouds}) by making a word cloud (in German---the English translation is in Table~\ref{tab:translation}) of the responses we received. To protect the privacy of our participants, we cannot report the verbatim German text.

\vspace{1mm}

\textbox{\textbf{Remark:} the lack of comments for screenshot \#1 (and its high agreement rating of \smamath{4.8}) is evidence that it deceived most users, but also shows that our questionnaire resembled a realistic phishing scenario wherein users are not primed.}

\subsection{Interpretation (with practitioners)}
\label{ssec:interpretation}
\noindent
Insofar, we have provided generic remarks that our participants expressed on these three screenshots. We now attempt to elaborate actionable insights---with the assistance of \companyname.

\textbf{Coding.} We held six meetings with \companyname's employees, focused on performing inductive coding sessions~\cite{thomas2006general}. The goal was to devise a codebook used to identify which (visual) elements in a screenshot of an AW can be used to infer that the corresponding image relates to a phishing webpage. Intuitively, by {\small \textit{(i)}}~identifying such elements, and then {\small \textit{(ii)}}~quantitatively measuring their prevalence ``in the wild'', it would be possible to determine which aspects should be prioritized by practitioners to improve the their PDS. During a meeting, the attendees discussed many screenshots of AW, attempting to derive an ``actionable'' set of phishing elements. After six meetings, the codebook encompasses 9 elements. Among these, we cite: ``altered visual logo'', ``different style of text and font'', and ``unusual login functionality and style''. 

\textbf{Mapping.}
We find it instructive to use the feedback received by our participants, and ``map'' it to these three abovementioned elements. This is useful as a form of validation: ``do netizens \textit{also} see the same elements that we see?''. Indeed, if these elements appear to have been noticed also by other internet users, then they can be acted upon to improve the detection mechanisms of PDS so that they can better deal with evasive phishing webpages. Due to the few comments received for screenshot \#1, we will only do this mapping for screenshot \#10 and \#18. We report in Table~\ref{tab:comments} our translation (which does not breach privacy) of those statements (in random order) that can be mapped to these three elements we identified. We recall that questions in part~III asked participants to explain ``why did you disagree with the statement in part~II?''. We could not find any statement for ``unusual login functionality'' (since \#10 does not have it in the first place---see Fig.~\ref{sfig:screenshot10}).

\begin{table*}[!htbp]
\caption{Mapping of participants' explanations (for Screenshots \#10 and \#18) to three of our codes. (Screenshot \#10 does not have any login form)}
\label{tab:comments}

\resizebox{2\columnwidth}{!}{

\begin{tabular}{c|l|l|l}
\toprule
  & \textbf{Altered Visual Logo} & \textbf{Unusual Login Functionality and Style} & \textbf{Different style of text and font}  \\
\midrule

\multirow{5}{*}{\rotatebox{90}{{\scriptsize \textbf{Screenshot \#10}}}}  
& ``because of the logo. It’s squeezed together'' & \multirow{5}{*}{N/A} & ``Looks a little distorted in the picture, not sure. May well be fake'' \\
& ``logo/branding looks fake. The font on the categories doesn’t fit.'' & & ``weird rendering and font'' \\
& ``Logo is not on top right and everything is very distorted/compressed'' & & ``Logo, Layout''\\
& ``Looks fake. (Logo, layout)'' & & ``The interface of Netflix looks different. The ”tabs” are arranged on the left, etc.'' \\
&  ``slightly different logo'' & &``Wasn’t exactly sure-the headings look different somehow (font \& size).'' \\

\midrule

\multirow{7}{*}{\rotatebox{90}{{\scriptsize \textbf{Screenshot \#18}}}}  

&``wrong Netflix logo - fake'' & ``Screenshot looks more like password renewa'' & ``modern login page looks different'' \\

&``wrong logo, it hasn't existed like this for years'' & ``completely different interface, Netflix doesn't use blue as much, generally different login and design'' & ``too minimalistic if you don't know the site'' \\

&``wrong logo'' & ``the Netflix login page looks different in my opinion'' & ``looks cheap, something is wrong there'' \\

& ``I find the logo weird, but it seems to be the page for registration, so not login but registration if the logo is not fake'' & ``you can see the registration page not the login page'' & ``Layout is too old fashioned, today Netflix login looks different'' \\

&``different logo and different colors'' & ``the login page looks different than what I'm used to. I find a little confusing/different'' & ``looks like a fake site'' \\

&``completely different logo'' & ``not login, but password change'' & ``outdated design'' \\

& & ``the registration page of Netflix that I know looks different'' &  \\

\bottomrule
\end{tabular}
}

\end{table*}

\begin{cooltextbox}
\textsc{\textbf{Takeaway.}}
Several participants noticed some ``common phishing elements'' that can be acted upon (by practitioners) to improve existing PDS against (real) evasive webpages.
\end{cooltextbox}

\textbf{Countermeasure.} Based on these explanations, \companyname is currently working towards a solution that is better equipped to counter similar phishing webpages---which can be somewhat detected by real users, but which are still an annoyance. Furthermore, an orthogonal objective pursued by \companyname is to identify some elements of AW that deceived most human users (e.g., \smamath{\#1, \#2, \#3}) and develop appropriate countermeasures. Nonetheless, we are currently working with \companyname to quantify the prevalence of ``elusive elements'' in AW, which can be used as a guide for practitioners to determine which elements are more common and hence should be given priority. 

\vspace{1mm}
\textbox{\textbf{``Would you change your mind?'':} Recall that our questionnaire ends with a binary question asking whether a given participant was willing to change their initial ratings if given another opportunity (§\ref{ssec:questionnaire}). Out of our 126 participants, 92 (\smamath{73\%}) affirmed that they would \textit{not} change their mind, whereas 34 (\smamath{27\%}) stated otherwise.}

\section{Discussion}
\label{sec:discussion}
\noindent
We discuss some alternative ways to carry out our study (§\ref{ssec:alternative}). Then, as a final contribution of our paper, we draw actionable recommendations for related research (§\ref{ssec:recommendations}).

\subsection{Alternative formulations}
\label{ssec:alternative}
\noindent
We discuss four specific design choices of our questionnaire. 
\begin{itemize}[leftmargin=0.5cm]
    \item \textbf{Structure of the questionnaire.} A valid point (which was also raised during our pilot study) is that asking the users to explain their disagreement \textit{after} having completed part~II can lead to users ``forgetting'' why they disagreed with some statements. We acknowledge such a remark: however, we did this because our main focus is determining if users are tricked by AW -- which is addressed by part~II. Asking the users to provide an explanation immediately after answering could have increased their suspiciousness, leading to less realistic responses for part~II (which is our main focus) in favor of more details for part~III (which relates to a relevant, but ancillary problem).
    
    \item \textbf{Phrasing of the questions in part~II.} Among our priorities was to minimize the amount of priming,\footnote{Designing bias-free user-studies for phishing is an open problem~\cite{sharma2021impact,alsharnouby2015phishing}.} which is why we opted for a neutral (and, potentially, vague\footnote{We never use terms such as ``trust'', ``malicious'', ``legitimate'', ``phishing''.}) question to be asked in part~II. Of course, we could have asked ``do you think that this screenshot represents a legitimate webpage?'' (similarly to, e.g.,~\cite{alsharnouby2015phishing}): however, doing so would have led our participants to be suspicious of every webpage---which is \textit{not realistic} in a phishing context, given that phishing is successful when users do not expect it (which also explains why most employees get phished despite receiving proper education~\cite{blancaflor2021let, burns2019spear}).\footnote{An interesting question to ask at the very end of our questionnaire is ``Did you figure out that this questionnaire was about phishing awareness (and, if so, when)?'', which would have acted as additional validation.}
    
    \item \textbf{Absence of context.} In our study, users are not given any information about ``why'' they would land on a given webpage. For instance, in a real setting, a user may be shown a webpage after clicking on a link (received, e.g., via email or instant messaging\footnote{In our questionnaire, we provided screenshots as rendered by a desktop Web-browser, hence we cannot assess the impact of phishing on mobile devices (e.g., smartphones, tablets). We encourage future research to do so.}). We acknowledge that context can be an important source to determine whether a website is phishing or not; however, our design choice is \textit{appropriate to answer our RQ}, whose goal is to investigate the susceptibility of users to AW, i.e., phishing webpages that evaded a PDS. If a user becomes suspicious of a webpage ``because of context'' then it would be unfair to the PDS (which, ttbook, do not account for context---yet). Furthermore, users who are sufficiently alert to become suspicious due to context are also less likely to fall for phishing in the first place~\cite{baki2022sixteen}: hence, lack of context can be seen as a scenario in which users do not suspect anything---which are the most dangerous, from a phishing perspective.
    
    \item \textbf{Number of AW.} Our questionnaire entails 18 AW (which are fixed\footnote{A random ordering could have been useful to, e.g., ascertain whether the skepticism over time is truly caused by the natural progression of the exercise, or by the different difficulties of the screenshots. However, random ordering would have also prevented a fair comparison for other effects that were more important for the sake of our study (see §\ref{ssec:questionnaire}).} for every participant), but \companyname provided us with a much higher number. While we acknowledge that we could have included more AW in our study, we did not do so for two reasons. First, because adding more AW to our questionnaire would have \textit{increased its length}, thereby: decreasing the level of attention of each participant; increasing the suspiciousness of each participant for any additional question; and potentially discouraging more users to participate (while part II was took \smamath{\sim}5 minutes to complete, part III took \smamath{\sim}15 minutes). Second, because having each participant provide their opinion on a different set of AW would have prevented one from \textit{analyzing trends} about individual AW (such as, e.g., investigating which AW tricked most users, and trying to understand ``why''). 
\end{itemize}
Simply put, there are many ways in which our study could have been designed---each with its pros and cons. Our choices are driven by our primary goals, dictated by our main RQ.

\subsection{Recommendations for Research}
\label{ssec:recommendations}
\noindent
Let us coalesce all our findings and derive recommendations for researchers. First, we endorse ``technical papers'' on phishing website detection to embrace our overarching message: \textit{carry out user-studies that focus on investigating how real users perceive the corresponding phishing websites}.\footnote{For papers that propose ``novel attacks'' that bypass existing anti-phishing schemes, such user-studies should verify whether users are really deceived by the evasive webpages; whereas papers that propose ``novel defenses'', the focus should be on the webpages that still mange to evade the robust PDS.} Then, we make three observations, rooted on our own experience and findings, that can help devising meaningful user-studies.
\begin{itemize}
    \item \textbf{It's feasible.} As our study showed, carrying out such user-studies is tough, but not impossible. Ultimately, we devised a questionnaire, advertised it on popular social media, and analysed the responses we collected over three weeks. Alternatively, the recent work by Lee et al.~\cite{lee2023attacking} relied on Amazon Mechanical Turk. Such an additional validation would dramatically increase the real-world value of the findings of a research paper.

    \item \textbf{Avoid priming.} Users are more skeptical of webpages when they are aware that they may be concealing a phishing trap---which may bias results. Hence, we recommend that future user-studies refrain from priming users. 
    
    \item \textbf{Make it short.} An important finding of our study is that, even when users are not primed, they may naturally become more suspicious of the samples shown during a questionnaire---if such samples exhibit strong elements that ``something is phishy.'' Hence, we recommend that {\small \textit{(i)}}~future user-studies only show few samples to any given participant, and that {\small \textit{(ii)}}~account for the fact that the responses for the last samples may be biased (due to the natural priming).
\end{itemize}
Finally, we remark that future efforts can even use our template (which we release~\cite{repository}) as basis for their questionnaires.
\section{Conclusions and Future Work}
\label{sec:conclusions}
\noindent
Countering phishing websites is a two-step decision process, entailing both ``machines'' (which provide a first layer of defense) and ``humans'' (who are the true target). Unfortunately, in research, prior work mostly focused on either one of these steps. At the same time, in reality, existing phishing detection systems (PDS) cannot detect all phishing websites, and end-users still fall for phish. 

In this paper, we advocate to \textit{change} the panorama of anti-phishing schemes in research. We do so by linking the response of humans with that of (real) machine learning-based PDS. We hope future endeavours will embrace the direction of our work. For instance, \textit{researchers} can assess the response of humans to their proposed PDS, thereby pinpointing which phishing techniques can simultaneously deceive both machines and users. The corresponding findings can then be used by \textit{practitioners} to refine their operational PDS. Ultimately, perfect detection is an enticing but unattainable goal: resources should be spent on countering those phishing webpages that are more likely to trick humans.

\textbf{\textsc{Ethical Statement}.} Our institutions are aware of and approve the research discussed in this paper. The respondents to our questionnaire know the identity of the author who collected their data, and we are willing to delete their data should they ask us to do so. The participants were made aware that their data was going to be privately stored, which is why we cannot disclose the full responses. To comply with the Menlo report~\cite{bailey2012menlo}, we never asked for sensitive data (i.e.,~\cite{sensitiveEU, sensitiveUSA}) or for personal identifiable information~\cite{krishnamurthy2009leakage} in our questionnaire; moreover, we only show screenshots of phishing webpages (which are reachable by unpublished links), and do not deploy any phishing webpage ``on the web''. We also complied with European regulation, and ensure that our participants are old enough to respond to online surveys without the explicit consent of a tutor (which varies from country to country~\cite{fra2020child}). Due to NDA with \companyname, we cannot disclose more information about the considered detector, the considered screenshots (which we do release in our repository~\cite{repository}), the effects that these screenshots had on \companyname's customers, or on \companyname itself.

\textbf{\textsc{Acknowledgements}}. The authors would like to thank the anonymous reviewers (as well as the eCrime'23 attendees) for their feedback. We also thank the Hilti Corporation for funding, and \companyname for allowing us to carry out this study.

\bibliographystyle{IEEEtran}

{\footnotesize


\begin{thebibliography}{10}
\providecommand{\url}[1]{#1}
\csname url@samestyle\endcsname
\providecommand{\newblock}{\relax}
\providecommand{\bibinfo}[2]{#2}
\providecommand{\BIBentrySTDinterwordspacing}{\spaceskip=0pt\relax}
\providecommand{\BIBentryALTinterwordstretchfactor}{4}
\providecommand{\BIBentryALTinterwordspacing}{\spaceskip=\fontdimen2\font plus
\BIBentryALTinterwordstretchfactor\fontdimen3\font minus
  \fontdimen4\font\relax}
\providecommand{\BIBforeignlanguage}[2]{{%
\expandafter\ifx\csname l@#1\endcsname\relax
\typeout{** WARNING: IEEEtran.bst: No hyphenation pattern has been}%
\typeout{** loaded for the language `#1'. Using the pattern for}%
\typeout{** the default language instead.}%
\else
\language=\csname l@#1\endcsname
\fi
#2}}
\providecommand{\BIBdecl}{\relax}
\BIBdecl

\bibitem{proofpoint2022phish}
``State of the phish 2022,''
  \url{https://www.proofpoint.com/it/resources/threat-reports/state-of-phish},
  ProofPoint, Tech. Rep., 2022.

\bibitem{khonji2013phishing}
M.~Khonji, Y.~Iraqi, and A.~Jones, ``Phishing detection: a literature survey,''
  \emph{IEEE Communications Surveys \& Tutorials}, 2013.

\bibitem{biggio2018wild}
B.~Biggio and F.~Roli, ``Wild patterns: Ten years after the rise of adversarial
  machine learning,'' \emph{Pattern Recognition}, 2018.

\bibitem{fbi2021icr}
``Interet crime report,''
  \url{https://www.ic3.gov/Media/PDF/AnnualReport/2022_IC3Report.pdf}, Federal
  Bur. of Investigation, Tech. Rep., 2022.

\bibitem{apwg2022}
``Phishing activity trends report,'' APWG, Tech. Rep., 2022,
  \url{https://docs.apwg.org/reports/apwg_trends_report_q2_2022.pdf}.

\bibitem{baki2022sixteen}
S.~Baki and R.~M. Verma, ``Sixteen years of phishing user studies: What have we
  learned?'' \emph{IEEE TDSC}, 2022.

\bibitem{orunsolu2018users}
A.~A. Orunsolu, O.~Afolabi, A.~S. Sodiya, A.~T. Akinwale \emph{et~al.}, ``A
  users' awareness study and influence of socio-demography perception of
  anti-phishing security tips,'' \emph{Acta Informatica Pragensia}, 2018.

\bibitem{abdelnabi2020visualphishnet}
S.~Abdelnabi, K.~Krombholz, and M.~Fritz, ``Visualphishnet: Zero-day phishing
  website detection by visual similarity,'' in \emph{ACM CCS}, 2020.

\bibitem{bell2020analysis}
S.~Bell and P.~Komisarczuk, ``An analysis of phishing blacklists: Google safe
  browsing, openphish, and phishtank,'' in \emph{ACSW}, 2020.

\bibitem{apruzzese2023real}
G.~Apruzzese, H.~S. Anderson, S.~Dambra, D.~Freeman, F.~Pierazzi, and
  K.~Roundy, ``{“Real Attackers Don't Compute Gradients”: Bridging the Gap
  Between Adversarial ML Research and Practice},'' in \emph{SaTML}, 2023.

\bibitem{oest2020phishtime}
A.~Oest, Y.~Safaei, P.~Zhang, B.~Wardman, K.~Tyers, Y.~Shoshitaishvili, and
  A.~Doup{\'e}, ``$\{$PhishTime$\}$: Continuous longitudinal measurement of the
  effectiveness of anti-phishing blacklists,'' in \emph{USENIX Security}, 2020.

\bibitem{divakaran2022phishing}
D.~M. Divakaran and A.~Oest, ``Phishing detection leveraging machine learning
  and deep learning: A review,'' \emph{IEEE Security \& Privacy}, 2022.

\bibitem{repository}
``{Our repo},'' \url{https://github.com/hihey54/eCrime23_realAdversarialPhish}.

\bibitem{ampel2023benchmarking}
B.~Ampel, Y.~Gao, J.~Hu, S.~Samtani, and H.~Chen, ``Benchmarking the robustness
  of phishing email detection systems,'' in \emph{ACIS}, 2023.

\bibitem{kersten2022investigating}
L.~Kersten, P.~Burda, L.~Allodi, and N.~Zannone, ``Investigating the effect of
  phishing believability on phishing reporting,'' in \emph{IEEE European
  Symposium on Security and Privacy Workshops}, 2022.

\bibitem{sheng2010falls}
S.~Sheng, M.~Holbrook, P.~Kumaraguru, L.~F. Cranor, and J.~Downs, ``Who falls
  for phish? a demographic analysis of phishing susceptibility and
  effectiveness of interventions,'' in \emph{Proc. Conf. HFCS}, 2010.

\bibitem{Corona:Deltaphish}
I.~Corona, B.~Biggio, M.~Contini, L.~Piras, R.~Corda, M.~Mereu, G.~Mureddu,
  D.~Ariu, and F.~Roli, ``Deltaphish: Detecting phishing webpages in
  compromised websites,'' in \emph{ESORICS}, 2017.

\bibitem{kondracki2021catching}
B.~Kondracki, B.~A. Azad, O.~Starov, and N.~Nikiforakis, ``Catching transparent
  phish: Analyzing and detecting mitm phishing toolkits,'' in \emph{ACM CCS},
  2021.

\bibitem{feal2021blocklist}
{\'A}.~Feal, P.~Vallina, J.~Gamba, S.~Pastrana, A.~Nappa, O.~Hohlfeld,
  N.~Vallina-Rodriguez, and J.~Tapiador, ``Blocklist babel: On the transparency
  and dynamics of open source blocklisting,'' \emph{IEEE Transactions on
  Network and Service Management}, 2021.

\bibitem{tian2018needle}
K.~Tian, S.~T. Jan, H.~Hu, D.~Yao, and G.~Wang, ``Needle in a haystack:
  Tracking down elite phishing domains in the wild,'' in \emph{IMC}, 2018.

\bibitem{zhang2007cantina}
Y.~Zhang, J.~I. Hong, and L.~F. Cranor, ``Cantina: a content-based approach to
  detecting phishing web sites,'' in \emph{Proc. WWW}, 2007.

\bibitem{apruzzese2022role}
G.~Apruzzese~et al, ``The role of machine learning in cybersecurity,''
  \emph{ACM Digital Threats: Research and Practice}, 2022.

\bibitem{mohammad2014predicting}
R.~M. Mohammad, F.~Thabtah, and L.~McCluskey, ``Predicting phishing websites
  based on self-structuring neural network,'' \emph{Neur. Comp. Appl.}, 2014.

\bibitem{cui2020semanticphish}
Q.~Cui, G.-V. Jourdan, G.~v. Bochmann, and I.-V. Onut, ``{SemanticPhish: a
  semantic-based scanning system for early detection of phishing attacks},'' in
  \emph{APWG Symposium on Electronic Crime Research (eCrime)}, 2020.

\bibitem{fu2006detecting}
A.~Y. Fu, L.~Wenyin, and X.~Deng, ``Detecting phishing web pages with visual
  similarity assessment based on earth mover's distance (emd),'' \emph{IEEE
  TDSC}, 2006.

\bibitem{van2021combining}
B.~Van~Dooremaal, P.~Burda, L.~Allodi, and N.~Zannone, ``Combining text and
  visual features to improve the identification of cloned webpages for early
  phishing detection,'' in \emph{Proc. ARES}, 2021.

\bibitem{lin2021phishpedia}
Y.~Lin, R.~Liu, D.~M. Divakaran \emph{et~al.}, ``Phishpedia: A hybrid deep
  learning based approach to visually identify phishing webpages,'' in
  \emph{Proc. USENIX Secur. Symp.}, 2021.

\bibitem{liu2022inferring}
R.~Liu, Y.~Lin, X.~Yang, S.~H. Ng, D.~M. Divakaran, and J.~S. Dong, ``Inferring
  phishing intention via webpage appearance and dynamics: A deep vision based
  approach,'' in \emph{USENIX Security}, 2022.

\bibitem{liang2016cracking}
B.~Liang \emph{et~al.}, ``Cracking classifiers for evasion: a case study on the
  google's phishing pages filter,'' in \emph{WWW}, 2016.

\bibitem{gao2023evading}
Y.~Gao, B.~M. Ampel, and S.~Samtani, ``Evading anti-phishing models: A field
  note documenting an experience in the machine learning security evasion
  competition 2022,'' \emph{ACM DTRAP}, 2023.

\bibitem{spacephish2022}
G.~Apruzzese, M.~Conti, and Y.~Yuan, ``Spacephish: The evasion-space of
  adversarial attacks against phishing website detectors using machine
  learning,'' in \emph{Proc. ACSAC}, 2022.

\bibitem{lee2023attacking}
J.~Lee, Z.~Xin, M.~P.~S. Ng, K.~Sabharwal, G.~Apruzzese, and D.~M. Divakaran,
  ``{Attacking logo-based phishing website detectors with adversarial
  perturbations},'' in \emph{European Symposium on Research in Computer
  Security (ESORICS)}, 2023.

\bibitem{hutchings2017online}
A.~Hutchings and T.~J. Holt, ``The online stolen data market: disruption and
  intervention approaches,'' \emph{Global Crime}, 2017.

\bibitem{zhang2022m}
P.~Zhang, Z.~Sun, S.~Kyung, H.~W. Behrens, Z.~L. Basque, H.~Cho, A.~Oest,
  R.~Wang, T.~Bao, Y.~Shoshitaishvili \emph{et~al.}, ``{I'm SPARTACUS, No, I'm
  SPARTACUS: Proactively Protecting Users from Phishing by Intentionally
  Triggering Cloaking Behavior},'' in \emph{Proc. ACSAC}, 2022.

\bibitem{apruzzese2022mitigating}
G.~Apruzzese and V.~Subrahmanian, ``Mitigating adversarial gray-box attacks
  against phishing detectors,'' \emph{IEEE Transactions on Dependable and
  Secure Computing}, 2022.

\bibitem{iuga2016baiting}
C.~Iuga, J.~R. Nurse, and A.~Erola, ``Baiting the hook: factors impacting
  susceptibility to phishing attacks,'' \emph{HCIS}, 2016.

\bibitem{lastdrager2017effective}
E.~Lastdrager, I.~C. Gallardo, P.~Hartel, and M.~Junger, ``How effective is
  anti-phishing training for children,'' in \emph{SOUPS}, 2017.

\bibitem{dhamija2006phishing}
R.~Dhamija, J.~D. Tygar, and M.~Hearst, ``Why phishing works,'' in \emph{Proc.
  SIGCHI CHI}, 2006.

\bibitem{tsow2007deceit}
A.~Tsow and M.~Jakobsson, ``Deceit and deception: A large user study of
  phishing,'' \emph{Indiana University.}, vol.~9, 2007.

\bibitem{sheng2007anti}
S.~Sheng \emph{et~al.}, ``Anti-phishing phil: the design and evaluation of a
  game that teaches people not to fall for phish,'' in \emph{SOUPS}, 2007.

\bibitem{jakobsson2007human}
M.~Jakobsson, ``The human factor in phishing,'' \emph{Privacy \& Security of
  Consumer Information}, vol.~7, no.~1, pp. 1--19, 2007.

\bibitem{herzberg2008security}
A.~Herzberg and A.~Jbara, ``Security and identification indicators for browsers
  against spoofing and phishing attacks,'' \emph{ACM TIT}, 2008.

\bibitem{alnajim2009anti}
A.~Alnajim and M.~Munro, ``An anti-phishing approach that uses training
  intervention for phishing websites detection,'' in \emph{IEEE ITNG}, 2009.

\bibitem{kumaraguru2010teaching}
P.~Kumaraguru, S.~Sheng, A.~Acquisti, L.~F. Cranor, and J.~Hong, ``Teaching
  johnny not to fall for phish,'' \emph{TOIT}, 2010.

\bibitem{yang2012building}
C.-C. Yang, S.-S. Tseng, T.-J. Lee, J.-F. Weng, and K.~Chen, ``Building an
  anti-phishing game to enhance network security literacy learning,'' in
  \emph{IEEE Int. Conf. Adv. Learn. Tech.}, 2012.

\bibitem{asanka2013can}
N.~Asanka, G.~Arachchilage, S.~Love, and C.~Maple, ``Can a mobile game teach
  computer users to thwart phishing attacks?'' \emph{International Journal for
  Infonomics}, 2013.

\bibitem{purkait2014empirical}
S.~Purkait, S.~Kumar~De, and D.~Suar, ``An empirical investigation of the
  factors that influence internet user’s ability to correctly identify a
  phishing website,'' \emph{Inf. Manag. \& Comp. Secur.}, 2014.

\bibitem{scott2014assessing}
M.~J. Scott, G.~Ghinea, and N.~A.~G. Arachchilage, ``Assessing the role of
  conceptual knowledge in an anti-phishing educational game,'' in \emph{IEEE
  Int. Conf. Advanced Learn. Tech.}, 2014.

\bibitem{alsharnouby2015phishing}
M.~Alsharnouby, F.~Alaca, and S.~Chiasson, ``Why phishing still works: User
  strategies for combating phishing attacks,'' \emph{IJHC}, 2015.

\bibitem{kunz2016nophish}
A.~Kunz, M.~Volkamer, S.~Stockhardt, S.~Palberg, T.~Lottermann, and E.~Piegert,
  ``Nophish: evaluation of a web application that teaches people being aware of
  phishing attacks,'' \emph{Informatik}, 2016.

\bibitem{arachchilage2016phishing}
N.~A.~G. Arachchilage, S.~Love, and K.~Beznosov, ``Phishing threat avoidance
  behaviour: An empirical investigation,'' \emph{Comp. Human Behavior}, 2016.

\bibitem{xiong2017domain}
A.~Xiong, R.~W. Proctor, W.~Yang, and N.~Li, ``Is domain highlighting actually
  helpful in identifying phishing web pages?'' \emph{Hum. Fact.}, 2017.

\bibitem{moreno2017fishing}
M.~M. Moreno-Fern{\'a}ndez, F.~Blanco, P.~Garaizar, and H.~Matute, ``Fishing
  for phishers. improving internet users' sensitivity to visual deception cues
  to prevent electronic fraud,'' \emph{Comp. Human Behav.}, 2017.

\bibitem{gopavaram2021cross}
S.~Gopavaram, J.~Dev, M.~Grobler, D.~Kim, S.~Das, and L.~J. Camp,
  ``Cross-national study on phishing resilience,'' in \emph{USEC}, 2021.

\bibitem{bullee2020effective}
J.-W. Bullee and M.~Junger, ``How effective are social engineering
  interventions? a meta-analysis,'' \emph{Information \& Computer Security},
  2020.

\bibitem{ferreira2015analysis}
A.~Ferreira and G.~Lenzini, ``An analysis of social engineering principles in
  effective phishing,'' in \emph{IEEE STAST Workshop}, 2015.

\bibitem{netflixStats}
D.~Ruby, ``40+ netflix statistics 2023.''
  \url{https://www.demandsage.com/netflix-subscribers/}, 2023, accessed March
  23, 2023.

\bibitem{amazonStats}
M.~Bożyczko, ``Amazon in europe: key statistics,''
  \url{https://nethansa.com/blog/amazon-in-europe-key-statistics}, 2023,
  accessed March 23, 2023.

\bibitem{zalandoStats}
Zalando, ``Zalando grows customer base and progresses on platform transition,''
  \url{https://corporate.zalando.com/en/financials/zalando-full-year-22-results},
  2023.

\bibitem{airbnbStats}
AirBnB, ``The eu host action plan,''
  \url{https://news.airbnb.com/wp-content/uploads/sites/4/2021/12/The-EU-Host-Action-Plan_Dec-2021.pdf},
  2023.

\bibitem{gmailStats}
FinancesOnline, ``Number of active gmail users: Statistics, demographics,
  usage.'' \url{https://financesonline.com/number-of-active-gmail-users/},
  2023.

\bibitem{googleStats}
M.~Mohsin, ``10 google search statistics you need to know in 2023,''
  \url{https://www.oberlo.com/blog/google-search-statistics}, 2023, accessed
  March 23, 2023.

\bibitem{instagramStats}
M.~Iqbal, ``Instagram revenue and usage statistics,''
  \url{https://www.businessofapps.com/data/instagram-statistics/}, 2023.

\bibitem{facebookStats}
Statista, ``Facebook monthly active users in europe as of 1st quarter 2023.''
  \url{https://www.statista.com/statistics/745400/facebook-europe-mau-by-quarter/},
  2023.

\bibitem{linkedinStats}
LinkedIn, ``Statistics---check out the numbers to understand the world’s
  largest professional network.'' \url{https://news.linkedin.com/about-us},
  2023.

\bibitem{paypalStats}
PayPal, ``Paypal announces nearly 35 million accounts in europe,''
  \url{https://newsroom.paypal-corp.com/2007-03-20-PayPal-Announces-Nearly-35-Million-Accounts-in-Europe},
  2023, accessed June 1, 2023.

\bibitem{uberStats}
Mappr, ``Countries with uber 2023,''
  \url{https://www.mappr.co/thematic-maps/countries-with-uber/}, 2023, accessed
  June 1, 2023.

\bibitem{yahooStats}
GitNux, ``The most surprising yahoo statistics and trends in 2023,''
  \url{https://blog.gitnux.com/yahoo-statistics/}, 2023, accessed June 1, 2023.

\bibitem{twitterStats}
DataReportal, ``Twitter users, stats, data and trends.''
  \url{https://datareportal.com/essential-twitter-stats}, 2023, accessed July
  1st, 2023.

\bibitem{horton2004qualitative}
J.~Horton, R.~Macve, and G.~Struyven, ``Qualitative research: experiences in
  using semi-structured interviews,'' in \emph{The real life guide to
  accounting research}.\hskip 1em plus 0.5em minus 0.4em\relax Elsevier, 2004.

\bibitem{krishnamurthy2009leakage}
B.~Krishnamurthy and C.~E. Wills, ``On the leakage of personally identifiable
  information via online social networks,'' in \emph{WOSN}, 2009.

\bibitem{fra2020child}
E.~A. for Fundamental~Rights, ``Child participation in research,''
  \url{https://fra.europa.eu/en/publication/2019/child-participation-research},
  2014.

\bibitem{alahmadi202299}
B.~A. Alahmadi, L.~Axon, and I.~Martinovic, ``99\% false positives: A
  qualitative study of $\{$SOC$\}$ analysts' perspectives on security alarms,''
  in \emph{Proc. USENIX Security Symp.}, 2022.

\bibitem{delacre2017psychologists}
M.~Delacre, D.~Lakens, and C.~Leys, ``Why psychologists should by default use
  welch's t-test instead of student's t-test,'' \emph{IRSP}, 2017.

\bibitem{lopez2022role}
P.~L{\'o}pez-Aguilar, C.~Patsakis, and A.~Solanas, ``The role of extraversion
  in phishing victimisation: A systematic literature review,'' in \emph{APWG
  Symposium on Electronic Crime Research (eCrime)}, 2022.

\bibitem{thomas2006general}
D.~R. Thomas, ``A general inductive approach for analyzing qualitative
  evaluation data,'' \emph{American journal of evaluation}, vol.~27, no.~2, pp.
  237--246, 2006.

\bibitem{sharma2021impact}
K.~Sharma, X.~Zhan, F.~F.-H. Nah, K.~Siau, and M.~X. Cheng, ``Impact of digital
  nudging on information security behavior: an experimental study on framing
  and priming in cybersecurity,'' \emph{OCJ}, 2021.

\bibitem{blancaflor2021let}
E.~B. Blancaflor, A.~B. Alfonso, K.~Banganay, G.~D. Cruz, K.~Fernandez, and
  S.~Santos, ``Let’s go phishing: A phishing awareness campaign using
  smishing, email phishing, and social media phishing tools,'' in \emph{IEOM},
  2021.

\bibitem{burns2019spear}
A.~Burns, M.~E. Johnson, and D.~D. Caputo, ``Spear phishing in a barrel:
  Insights from a targeted phishing campaign,'' \emph{JOCEC}, 2019.

\bibitem{bailey2012menlo}
M.~Bailey, D.~Dittrich, E.~Kenneally, and D.~Maughan, ``{The Menlo report},''
  \emph{IEEE Security \& Privacy}, 2012.

\bibitem{sensitiveEU}
E.~Commission, ``Sensitive data,''
  \url{https://ec.europa.eu/info/law/law-topic/data-protection/reform/rules-business-and-organisations/legal-grounds-processing-data/sensitive-data_en}.

\bibitem{sensitiveUSA}
U.~D. of~the Treasury, ``Sensitive personal data,''
  \url{https://home.treasury.gov/taxonomy/term/7651}.

\end{thebibliography}
}

\appendix
\section{Appendix: Tables and Figures}
\label{app:first_appendix}

\begin{figure}[!htbp]
    \centering
    \begin{subfigure}[!htbp]{0.99\columnwidth}
        \centering
        \includegraphics[width=1\columnwidth]{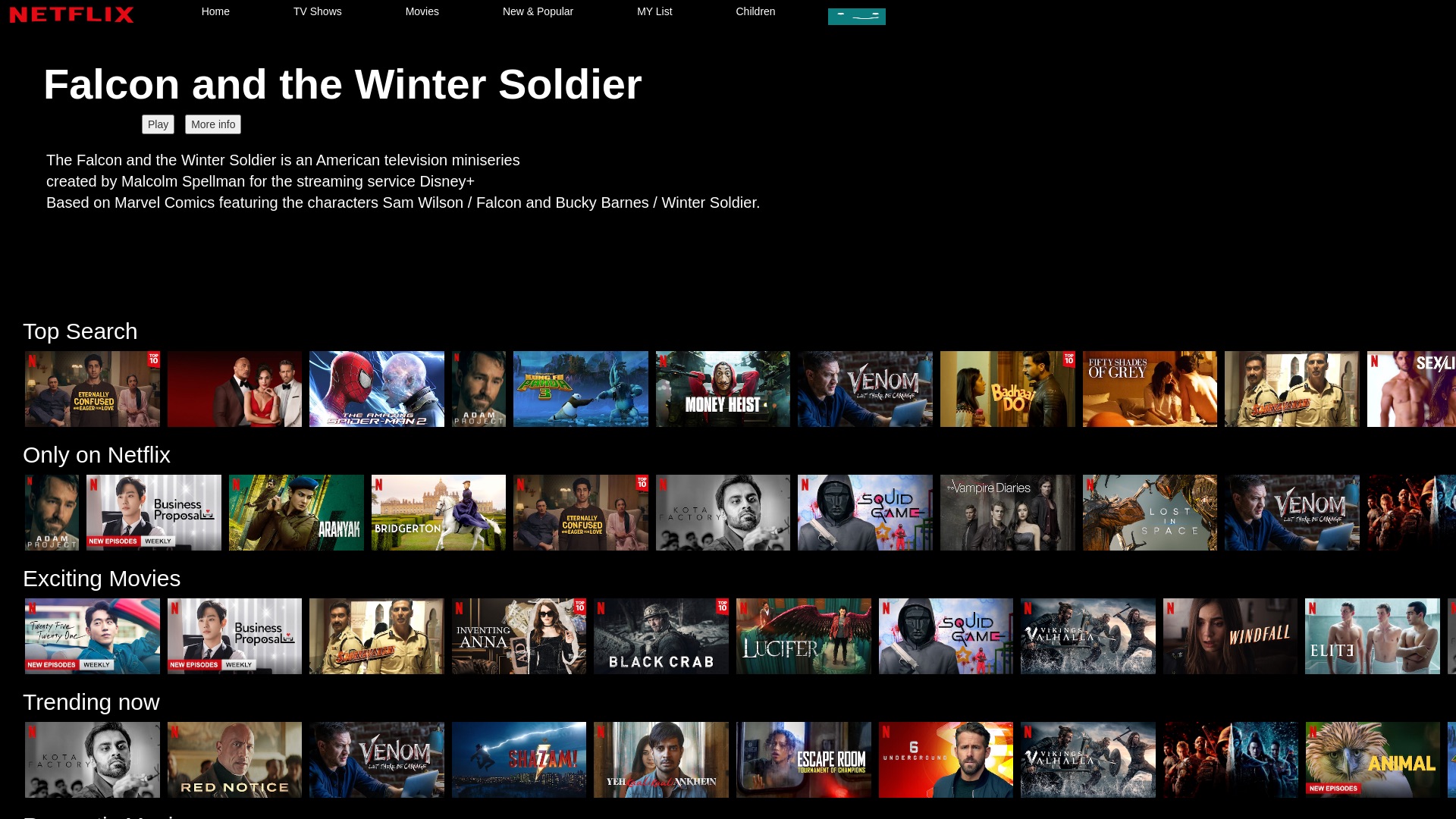}
        \caption{Screenshot 10 (``moderate difficulty'' to identify as phishing---by humans).}
        \label{sfig:screenshot10}
    \end{subfigure}

\vspace{2mm}
    
    \begin{subfigure}[!htbp]{0.99\columnwidth}
        \centering
        \includegraphics[width=1\columnwidth]{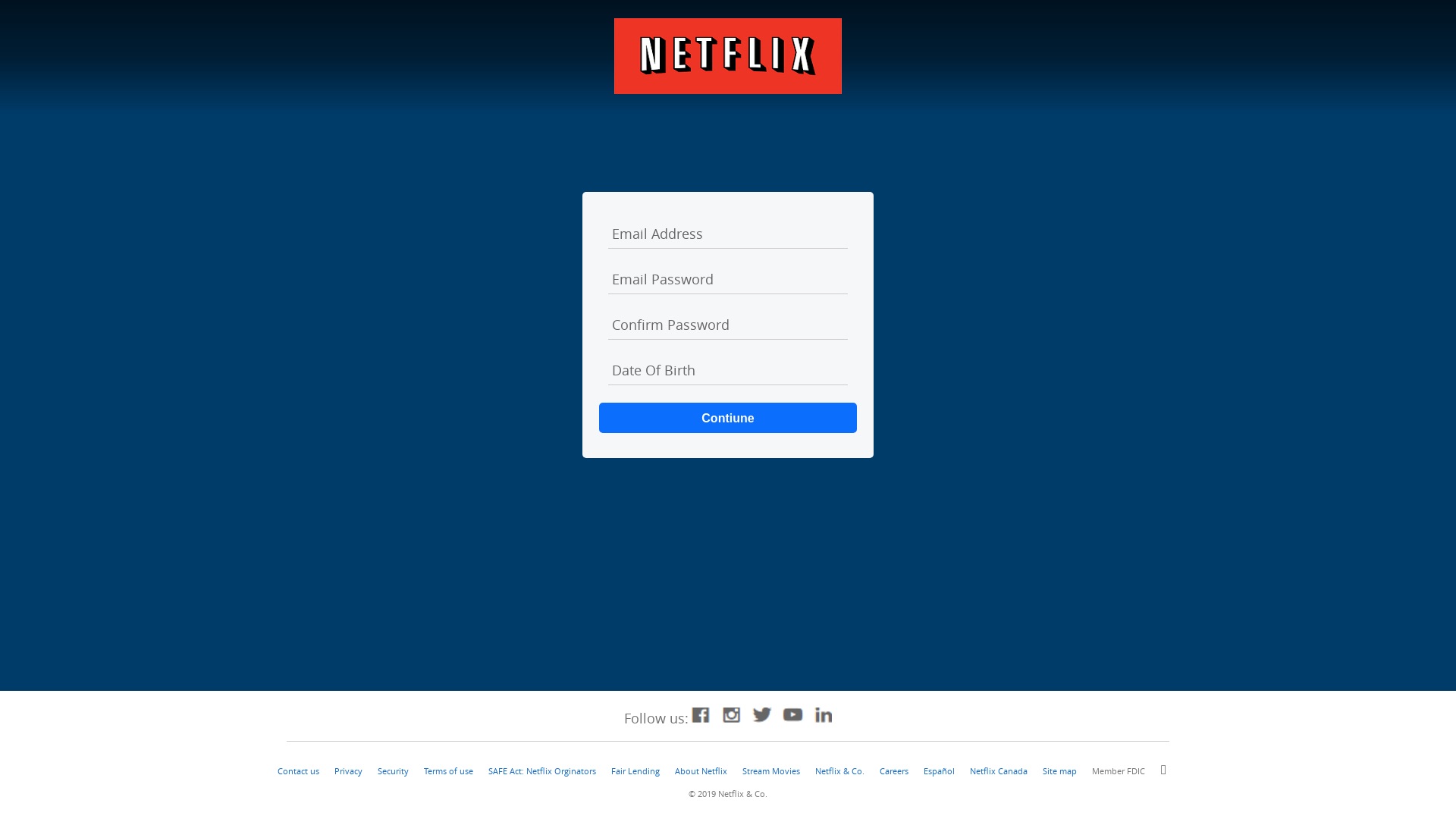}
        \caption{Screenshot 18 (``very easy difficulty'' to identify as phishing---by humans).}
        \label{sfig:screenshot18}
    \end{subfigure}
    \caption{Two screenshots of adversarial phishing webpages (mimicing NetFlix) included in our questionnaire.}
    \label{fig:screenshots}
\end{figure}

\begin{figure}[!htbp]
    \centering
    \begin{subfigure}[!htbp]{0.49\columnwidth}
        \centering
        \includegraphics[width=1\columnwidth]{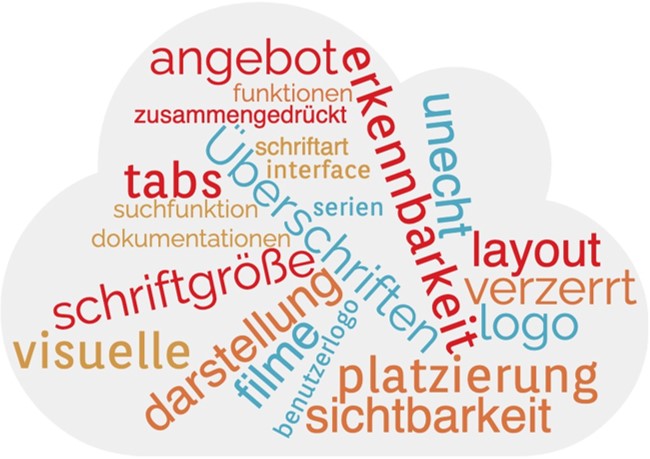}
        \caption{Comments on AW \#10 (Fig.~\ref{sfig:screenshot10}).}
        \label{fig:wordcloud10}
    \end{subfigure}
    \begin{subfigure}[!htbp]{0.49\columnwidth}
        \centering
        \includegraphics[width=1\columnwidth]{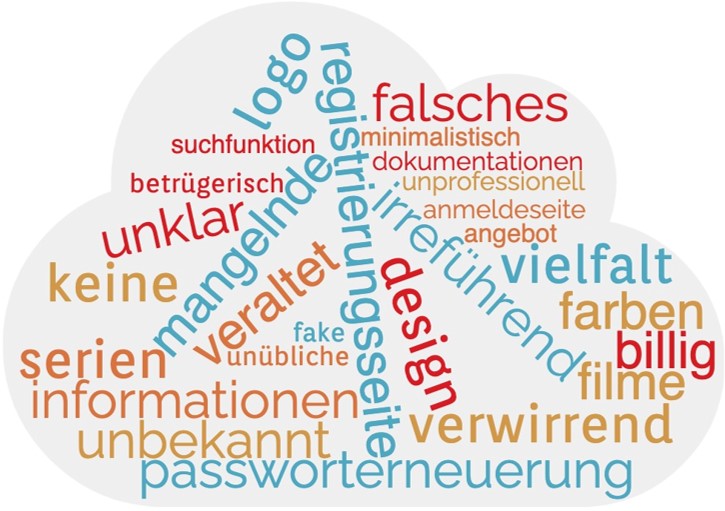}
        \caption{Comments on AW \#18 (Fig.~\ref{sfig:screenshot18}).}
        \label{sfig:wordcloud18}
    \end{subfigure}
    \caption{Word clouds (in German) for two screenshots (both mimicing NetFlix). Refer to Table~\ref{tab:translation} for the English translation of the terms.}
    \label{fig:wordclouds}
\end{figure}

\begin{table}[!htbp]
    \centering
    \caption{English translation of the word cloud }
    \label{tab:translation}

        \begin{subtable}[h]{0.45\textwidth}
            \centering
            \caption{Screenshot 10 (AW) in Fig.~\ref{fig:wordcloud10}}
            \label{stab:translation_10}
            \begin{tabular}{l|l}
                \toprule
                \textbf{German} & \textbf{English} \\
                \midrule
                Logo & Logo\\
                Verzerrt & Distorted \\
                Zusammengedrückt & Compressed\\
                Unecht & Unreal \\
                Layout & Layout\\
                Schriftart & Font\\
                Filme & Movies \\
                Serien & Series\\
                Dokumentationen & Documentaries\\
                Interface & Interface\\
                Tabs & Tabs \\
                Suchfunktion & Search function\\
                Benutzerlogo & User logo\\
                Überschriften & Headings\\
                Sichtbarkeit & Visibility \\
                Funktionen & Functions\\
                Angebot & Offer \\
                Visuelle Darstellung & Visual Presentation\\
                Platzierung & Placement \\
                Schriftgröße & Font size \\
                Erkennbarkeit & Recognizability \\
                \bottomrule
            \end{tabular}
        \end{subtable}
        
        \vspace{4mm}
        
        \begin{subtable}[h]{0.45\textwidth}
            \centering
            \caption{Screenshot 18 (AW) in Fig.~\ref{sfig:wordcloud18}}
            \label{stab:translation_18}
            
            \begin{tabular}{l|l}
                \toprule
                \textbf{German} & \textbf{English} \\
                \midrule
                Falsches Logo & Wrong Logo \\
                Veraltet & Outdated \\
                Unprofessionell & Unprofessional \\
                Design & Design \\
                Mangelnde Informationen & Lack of Information \\
                Keine Suchfunktion & No search function \\
                Passworterneuerung & Password Renewal\\
                Unübliche Farben & Unusual colors\\
                Unklar & Unclear \\
                Fake & Fake \\
                Minimalistisch & Minimalistic \\
                Unbekannt & Unknown \\
                Betrügerisch & Fraudolent \\
                Registrierungsseite & Registration page\\
                Anmeldeseite & Registration page \\
                Verwirrend & Confusing \\
                Billig & Cheap \\
                Irreführend & Misleading \\
                Serien & Series \\
                Filme & Movies \\
                Dokumentationen & Documentaries \\
                Vielfalt & Variety \\
                Angebot & Offer \\
                \bottomrule
            \end{tabular}
        \end{subtable}

\end{table}

\end{document}